\documentclass[]{emulateapj}

\usepackage{wasysym} 
\usepackage{amssymb}
\usepackage{lscape}

\begin{document}

\title{Are Optically-Selected Quasars Universally X-Ray Luminous?  X-Ray/UV Relations in Sloan Digital Sky Survey Quasars}

\author{Robert R. Gibson, W. N. Brandt, Donald P. Schneider}
\affil{Department of Astronomy and Astrophysics\\Pennsylvania State University\\525 Davey Laboratory\\University Park, PA, 16802}
\email{rgibson@astro.psu.edu}

\shorttitle{X-ray Emission from SDSS Non-BAL QSOs}
\shortauthors{Gibson et al.}

\journalinfo{Accepted to ApJ. (c) Copyright 2008.  The American Astronomical Society.  All rights reserved.  Printed in U.S.A.}
\hbadness=10000

\clearpage

\begin{abstract}
We analyze archived {\it Chandra} and {\it XMM-Newton} \mbox{X-ray} observations of 536 Sloan Digital Sky Survey (SDSS) Data Release 5 (DR5) quasars (QSOs) at $1.7 \le z \le 2.7$ in order to characterize the relative UV and \mbox{X-ray} spectral properties of QSOs that do not have broad UV absorption lines (BALs).  We constrain the fraction of \mbox{X-ray} weak, non-BAL QSOs and find that such objects are rare; for example, sources underluminous by a factor of 10 comprise $\la$2\% of optically-selected SDSS QSOs.  \mbox{X-ray} luminosities vary with respect to UV emission by a factor of $\la$2 over several years for most sources.  UV continuum reddening and the presence of narrow-line absorbing systems are not strongly associated with \mbox{X-ray} weakness in our sample.  \mbox{X-ray} brightness is significantly correlated with UV emission line properties, so that relatively \mbox{X-ray} weak, non-BAL QSOs generally have weaker, blueshifted \ion{C}{4}~$\lambda$1549 emission and broader \ion{C}{3}]~$\lambda$1909 lines.  The \ion{C}{4} emission line strength depends on both UV and \mbox{X-ray} luminosity, suggesting that the physical mechanism driving the global Baldwin effect is also associated with \mbox{X-ray} emission.

\end{abstract}


\section{INTRODUCTION\label{introSec}}

It has been known for some time that the UV and \mbox{X-ray} luminosities of quasars (QSOs) are correlated \citep[e.g.,][and references therein]{at82}, and recent studies have carefully quantified this relation across $\approx$5 orders of magnitude in UV luminosity \citep[e.g.,][]{sbsvv05, ssbaklsv06, jbssscg07}.  Such studies inform ongoing efforts to understand the structure and physics of QSO nuclear regions, providing quantitative constraints on models of physical associations between UV and \mbox{X-ray} emission.  Because UV photons are generally believed to be radiated from the QSO accretion disk while \mbox{X-rays} are produced in the disk corona \cite[e.g.,][and references therein]{rn03}, the \mbox{UV/X-ray} luminosity relation is an indication of the balance between accretion disks and their coronae.  For example, a large fraction of intrinsically X-ray weak sources would suggest that coronae may frequently be absent or disrupted in QSOs.  In this study, we combine results from recent optical/UV and \mbox{X-ray} observations of hundreds of QSOs in order to constrain the fraction of sources that are anomalously \mbox{X-ray} weak, and also to test for additional physical effects that contribute to the scatter observed in the relation between UV and \mbox{X-ray} luminosities.

Radio-loud QSOs are well known to be relatively \mbox{X-ray} bright \citep[e.g.,][]{wtgz87, blvsbbwg00}, while QSOs with broad UV absorption lines (BALs) are \mbox{X-ray} faint \citep[e.g.,][]{gsahfbftm95, blw00}.  We wish to understand the emission processes of QSOs without the added complexity of the strong X-ray absorption associated with UV BAL outflows or the enhanced X-ray emission that may be associated with a radio jet.  In order to accomplish this, BAL and radio-loud QSOs must be carefully removed from samples.  This process can be complicated in cases where available spectra do not extend to wavelengths where strong UV BAL absorption may occur.

In this study, we compare the UV and \mbox{X-ray} properties of optically-selected, radio-quiet Sloan Digital Sky Survey \citep[SDSS; e.g.,][]{y+00} QSOs that are known not to host BAL outflows along the line of sight.  Careful screening enables us to study the ways in which these ``ordinary'' QSOs deviate from the general relation between UV and \mbox{X-ray} luminosities.  We quantify the scatter about the best-fit relation between UV and \mbox{X-ray} luminosities, and in particular constrain the fraction of optically-selected QSOs that are intrinsically weak \mbox{X-ray} emitters.  Secondly, we search for correlations between UV emission/absorption properties and {\it relative} \mbox{X-ray} brightness, where the term ``relative'' indicates the observed \mbox{X-ray} brightness compared to that expected for an average QSO with the same UV luminosity.  Beyond the UV/X-ray luminosity relation, our findings place additional constraints on physical models that relate QSO UV and \mbox{X-ray} emission processes.

The significant amount of scatter in the \mbox{X-ray} brightness of individual sources compared to that of ``average'' QSOs with the same UV luminosity \citep[e.g.,][]{ssbaklsv06} indicates that some unmodeled physical factors influence the relation between UV and \mbox{X-ray} luminosities.  Besides the cases of radio-loud and BAL QSOs mentioned above, possible additional causes of scatter include additional \mbox{X-ray} absorption that is not associated with UV BALs, intrinsically weak (or strong) \mbox{X-ray} emission, and time variability.  Studies of \mbox{X-ray} brightness in QSOs use different methods to quantify UV absorption and \mbox{X-ray} weakness (complicating comparisons between different samples), and intensive UV and \mbox{X-ray} spectroscopic campaigns have been performed for only a few QSOs.  As a result, we do not have strong constraints on how frequently \mbox{X-ray} absorption, intrinsically faint \mbox{X-ray} emission, and time variability cause \mbox{X-ray} weakness in QSO samples.  In the remainder of this section, we briefly review published cases involving these physical scenarios.  This discussion includes all the non-BAL QSOs listed in \citet{blw00} and five of the six sources from \citet{ybsv98} with $\alpha_{OX} < -2$.\footnote{The sixth source, QSO~$0316-346$, was identified from optical spectra by \citet{mppb88}.  UV spectra are required to search for BAL absorption, but no UV spectra of this source are present in the MAST or HEASARC archives.}  We also discuss cases of unusual X-ray weakness in recent studies of individual sources.

\subsection{X-Ray Absorption\label{xAbsCauseXWSec}}

Even for QSOs that do not show UV BAL absorption, strong \mbox{X-ray} absorption may be the simplest explanation for \mbox{X-ray} weakness.  Mrk~304 (PG~$2214+139$) is an \mbox{X-ray} weak source that is often classified as a Seyfert~1 galaxy.  It does not show UV BAL absorption, although \citet{lb02} found some evidence for moderate \ion{C}{4} and \ion{N}{5} line absorption.  In the \mbox{X-rays}, it is strongly absorbed by a multizone, ionized absorber with a total column density $N_H \sim 10^{23}$~cm$^{-2}$ \citep{pjgsrs04, bpf04}.  Another \mbox{X-ray} weak source, PG~$1126-041$, shows evidence for an ionized \mbox{X-ray} absorber.  However, low-velocity, broad \ion{C}{4} absorption has been found in some observations, indicating that PG~$1126-041$ should be classified as a (variable) BAL QSO \citep{wbwyw99}.

\subsection{Intrinsic X-Ray Weakness\label{intrinsicCauseXWSec}}

\mbox{X-ray} weakness cannot always be attributed to absorption.  The nearby ($z = 0.192$), narrow-line type 1 QSO PHL~1811 shows no evidence of BALs in its UV spectrum.  With an upper limit for a neutral \mbox{X-ray} absorbing column of $N_H < 8.7 \times 10^{20}$~cm$^{-2}$, the source is not strongly \mbox{X-ray} absorbed \citep{lhjgcp07}.  It has been consistently \mbox{X-ray} weak since it was first observed almost 20 years ago with {\it ROSAT}, although the \mbox{X-ray} flux has varied by a factor $\approx$5.  This variation (small compared to the degree of \mbox{X-ray} weakness) suggested to \citet{lhjgcp07} that the \mbox{X-ray} emission was not scattered into view, as the medium responsible for the scattering would need to be implausibly small.  PHL~1811 therefore appears to be an intrinsically weak \mbox{X-ray} emitter.  The optical/UV spectrum of PHL~1811 is also unusual \citep{lhjc07}.  It is very blue and shows no forbidden or semi-forbidden line emission.  The \ion{C}{4} $\lambda 1549$ line emission is weak by a factor of $\approx5$ compared to the composite spectrum of \citet{fhfcwm91}.  In this study, we will test whether intrinsically \mbox{X-ray} weak objects like PHL~1811 (which would have been flagged for SDSS spectroscopy based on its optical colors alone) are common, and whether these unusual emission line characteristics are associated with \mbox{X-ray} weakness.

Another source, PG~$1011-040$, is relatively \mbox{X-ray} weak by a factor of $\sim$10 \citep{blw00, gblemwi01}.  There is some evidence for \ion{C}{4} absorption in the UV spectrum, but no strong indication of a BAL \citep{blw00, lb02}.  It appears to be relatively unabsorbed in \mbox{X-rays}, with an upper limit on the absorbing column density of $N_H \le 5\times 10^{21}$~cm$^{-2}$.  In contrast to PHL~1811, multiple observations of PG~$1011-040$ have not found much \mbox{X-ray} luminosity variation \citep{gblemwi01}.

\subsection{UV and X-Ray Variability\label{varCauseXWSec}}

Source variability may be an important factor contributing to the scatter in the UV/\mbox{X-ray} luminosity relation.  Additional scatter is introduced by the fact that optical/UV and \mbox{X-ray} observations are usually taken some time apart, and this time may reach up to years.

The narrow-line Seyfert~1 galaxy Mrk~335 has recently been reported to have decreased in \mbox{X-ray} brightness by a factor of $\approx$30 \citep{gkg07}.  The authors suggested the variation may have been caused by the onset of heavy \mbox{X-ray} absorption, which may also lead to the appearance of UV BALs.  PG~$0844+349$ has been observed to vary in the \mbox{X-rays} by up to 60\% on short (20~ks) time scales, and by a factor of $\sim$10 on longer (multi-year) time scales \citep[e.g.,][]{gblemwi01, bgbf03}.  The narrow-line Seyfert 1 galaxy WPVS~007 has dimmed in the \mbox{X-rays} by a factor of 100 or more, while a \ion{C}{4} BAL appears to be forming in the UV spectrum of this source \citep{gslkon07}.

Finally, we note that our understanding of the causes of \mbox{X-ray} weakness is limited by observing time scales.  Over longer time scales, \mbox{X-ray} weak QSOs may show different properties (such as long-term emission or absorption variation), and may even cease to be \mbox{X-ray} weak.  We have also noted in \S\ref{introSec} cases where UV BALs may be transient, and recent study has demonstrated that BALs evolve on multi-year (rest-frame) time scales \citep[e.g.,][]{gbsg08}.  For these reasons, we are only able to place limits on the extent of \mbox{UV/X-ray} variability (\S\ref{uVXVarConstraintSec}).  Future near-simultaneous, multi-wavelength observations of QSOs will improve on these constraints.

Throughout this work we use a cosmology in which $H_0 = 70$~km~s$^{-1}$~Mpc$^{-1}$, $\Omega_M = 0.3,$ and $\Omega_{\Lambda} = 0.7$.

\section{OBSERVATIONS AND DATA REDUCTION\label{obsDataSec}}

The SDSS Data Release 5 \citep[DR5;][]{a-m+07} has recently cataloged spectra of over 77,000 QSOs, most of which were selected for spectroscopy based on optical considerations \citep{s+07}.  Many of these sources have also been observed in the \mbox{X-rays} with modern telescopes such as {\it Chandra} and {\it XMM-Newton}, enabling detailed studies of relations between UV and \mbox{X-ray} properties for a large sample of optically-selected QSOs.  

In this study, we consider the 536 SDSS QSOs at redshifts $1.7 \le z \le 2.7$ that have been observed with {\it Chandra} or {\it XMM-Newton}.  The lower limit on redshift ensures that the SDSS spectrum reaches to $\approx$1400~\AA, allowing good coverage of any potential high-velocity \ion{C}{4} absorption, up to $\approx-30,000$~km~s$^{-1}$.  The upper redshift limit ensures that the 2500~\AA\ continuum is in view so that we can determine the UV continuum flux accurately.  We obtain QSO redshifts from the SDSS DR5 QSO catalog \citep{s+07}.  These redshifts were generated by the SDSS spectroscopic pipeline based on the overall continuum and emission properties of each spectrum; a small fraction ($\approx$1\%) were revised during visual inspection for the QSO catalog.  The mean error in redshift estimated from the pipeline for our sources is $\approx$0.002.  An rms difference of 0.006 in redshift determination has been observed in sources observed multiple times in the SDSS \citep{wvkspbrb05, s+07}.  We therefore expect that errors in redshift are typically 0.5\% or less for our sources.  We note recent studies have found no significant evidence for redshift evolution in $\alpha_{OX}$ \citep[e.g.,][]{ssbaklsv06}, X-ray spectral shape \citep[e.g.,][]{sbvsfrs05}, or infrared continuum properties \citep[e.g.,][]{jfhsvbbcclprrsswb06}, so we do not expect that our results are strongly influenced by our redshift restriction.

\begin{figure} [ht]
  \begin{center}
      \includegraphics[width=4in, angle=270]{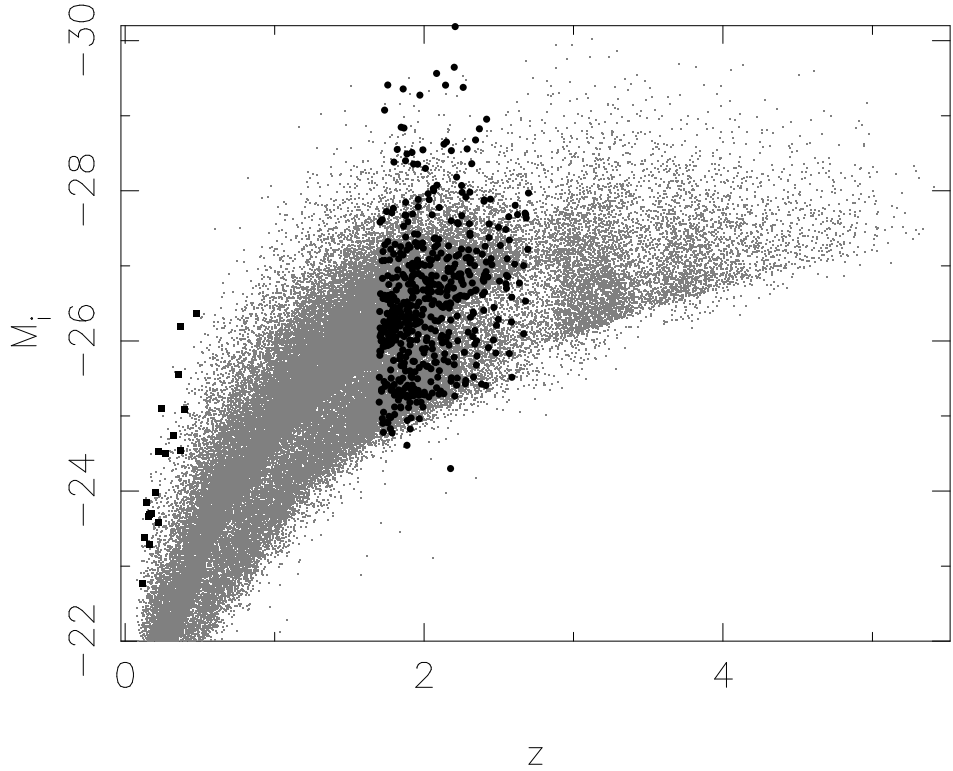}
      \caption{\label{sDSSBQSPlotFig}The redshifts and absolute $i$ magnitudes, $M_i$, for our 536 sample $A$ sources at $1.7 \le z \le 2.7$ (filled circles) compared to the entire SDSS DR5 QSO catalog (gray points).  We also indicate (with filled squares) the BQS QSOs at $z < 0.5$ which were included in the SDSS DR5 QSO catalog.}
   \end{center}
\end{figure}

In order to illustrate the parameter space covered by our sample, we plot in Figure~\ref{sDSSBQSPlotFig} the redshifts and absolute $i$ magnitudes for our sources together with the full DR5 QSO catalog.  For comparison, we also indicate SDSS DR5 QSOs at $z < 0.5$ which were included in the Bright Quasar Survey \citep[BQS;][]{sg83} in the plot.

\subsection{X-Ray Observation Catalog\label{obsCatSec}}

We obtained the list of all {\it Chandra} observations that were publicly available as of 2007 July 26 and searched for cases where a DR5 QSO fell within 15\arcmin~of the telescope pointing for an ACIS observation with no gratings.  For this subset of {\it Chandra} observations, we tested whether each QSO was on an ACIS chip using the CIAO tool {\tt dmcoords}.  We retrieved observations with on-chip sources from the {\it Chandra} online database and ran {\tt acis\_process\_events} on each data set in order to apply the latest calibration data, including the time-dependent gain correction and charge transfer inefficiency (CTI) corrections.\footnote{See the CIAO threads http://asc.harvard.edu/ciao/threads/acistimegain/and http://asc.harvard.edu/ciao/threads/acisapplycti/}  All processing was performed using CIAO version 3.4.\footnote{http://asc.harvard.edu/ciao3.4/}

We extracted spectra for the source and the background of each on-chip DR5 QSO using the CIAO tool {\tt psextract}.  For the source extraction region, we used a circle with radius equal to the 90\% encircled energy fraction plus an additional 5 pixels.  For the background region, we used an annulus around the source with inner (outer) radii equal to the 90\% encircled energy fraction plus 15~(50) pixels.  We chose these regions because we have found that they accurately characterize the surrounding \mbox{X-ray} background with a minimum of contamination by nearby sources in the background regions.  Because a large fraction of our sources are serendipitously observed at large off-axis angles and may be faint in any case, we are concerned to avoid this contamination.

Because the CIAO tool {\tt psextract} does not generate spectra for sources with zero counts, we flagged these cases for proper handling by our fitting algorithm described in \S\ref{fitXRaySec}.  We also excluded observations where the source lies within 32 pixels of a chip edge.

We used {\tt wavdetect} \citep{fkrl02} with the {\tt sigthresh} parameter set to $10^{-5}$ to detect any sources that may contaminate the annular regions we used to extract background spectra.  In six cases, we found that the total number of counts contributed to the background spectrum by detected sources was $\ge10$\% of the observed spectrum.  In these cases, we manually adjusted the background extraction regions to be free of source contamination and re-processed the data.

We also obtained a list of all {\it XMM-Newton} observations that were public as of 2007 August 22 for which a DR5 QSO lay within 15\arcmin~of the target.  We reduced EPIC camera data using the {\tt emchain} and {\tt epchain} tasks in the SAS~7.1.0 package.\footnote{http://xmm.vilspa.esa.es/sas/}  We removed intervals of background flaring from each observation by flagging all time intervals for which the count rate exceeded the baseline count rate by a factor of three or more, where the baseline count rate was defined as the rate for which the distribution of count rates peaked.

We extracted spectra for the MOS and $pn$ cameras using a circular aperture 15\arcsec~in radius for the source and an annulus with inner~(outer) radii of 50\arcsec~(75\arcsec).  For each arcminute that the source was off-axis from the nominal pointing, we increased the extraction radii by 1\arcsec~(2.3\arcsec) for the MOS~($pn$) cameras to account for the increase in the constant encircled energy fraction radius.  As for the {\it Chandra} processing, the extraction regions were chosen to characterize effectively the surrounding background while minimizing contamination from nearby sources.  We used the SAS utility {\tt eboxdetect} to detect sources that may contaminate our background regions.  In cases where bright sources were detected in our background extraction regions, we omitted the the brightest source from the background extraction in order to improve background estimation.

A single {\it XMM-Newton} observation may include one or more exposures from the MOS1, MOS2, and $pn$ cameras.  From the available set of exposures, we preferentially selected the MOS observation with the longest exposure.  Although the $pn$ cameras provide a larger effective area, we found that the higher angular resolution of the MOS cameras yield more reliable background estimation and a higher fraction of source detections.  For this reason, we prefer the MOS cameras for our study.

We discard sources for which the effective area is $<$100~cm$^{2}$ in all bins in order to filter out sources that are near a chip edge.  Even so, we find that the fraction of undetected sources is much greater in the {\it XMM-Newton} reductions than for {\it Chandra}, due to the significantly better angular resolution of the {\it Chandra} telescope.  In cases where undetected sources are important for our statistics, we exclude the {\it XMM-Newton} observations from consideration.

In eight cases, the number of background counts attributed to {\tt eboxdetect} sources in the background extraction regions was $>$30\% of the total counts extracted from the source region.  Visual inspection indicates that most of these are cases where the source was weak and the candidate background sources are borderline detections.  We manually selected a suitable background and re-processed the two cases where strong sources clearly impact the background estimation.

Approximately 9\% of our {\it Chandra} QSOs and 1\% of our {\it XMM-Newton} QSOs were targets of observations.  (The fraction is higher for {\it Chandra} because targets observed with both telescopes will only have their {\it Chandra} observations included in our sample.)  While there may be small biases introduced by the tendency of observers to select unusual objects for observation, our study is dominated by sources observed serendipitously.  We remove targeted sources from our sample when appropriate.

In all, we have \mbox{X-ray} observations of 536 SDSS QSOs at $1.7 \le z \le 2.7$.  Of these, 315 were obtained with {\it Chandra}, while 221 were obtained with the {\it XMM-Newton} MOS cameras.  We considered any source that was detected (using Poisson statistics) at 99\% confidence above the background in the soft (0.5--2~keV), hard (2--8~keV), or full (0.5--8~keV) bands to be detected in \mbox{X-rays}.  90\% of the {\it Chandra} sources and 67\% of the {\it XMM-Newton} sources in our sample were detected by this criterion.  We provide a catalog of our \mbox{X-ray} observations in Table~\ref{sourceInfoTab}.

\subsection{Fitting SDSS Spectra\label{fitSDSSSec}}

Before fitting the SDSS spectra, we multiply them by a constant to match (approximately) the $g$, $r$, and $i$ PSF magnitudes synthesized from the spectra to the measured photometric magnitudes.  We also correct the spectra for Galactic extinction using the reddening curve of \citet{ccm89} with the near-UV extension of \citet{o94}.  We obtain $E(B-V)$ from the NASA Extragalactic Database (NED)\footnote{http://nedwww.ipac.caltech.edu/}, which uses the dust maps of \citet{sfd98}.  The QSOs in our study are at such high redshifts and luminosities that their spectra are not significantly contaminated by their host galaxies.

We fit SDSS spectra using the algorithm of \citet{gbsg08}, which we summarize here.  Our continuum model is a power law reddened using the Small Magellanic Cloud reddening curve of \citet{p92}.  We do not attach physical significance to the intrinsic $E(B-V)$ values we obtain from these fits, as the reddening is degenerate with the shape of the underlying continuum emission.  The UV luminosities we report are therefore corrected for Galactic, but not intrinsic, reddening.  This follows the general practice for studies of \mbox{UV/X-ray} luminosity relations.  We fit regions that are generally free from strong absorption or emission features (1250--1350, 1600--1800, 1950--2050, 2150--2250, and 2950--3700~\AA), and then re-fit the spectrum iteratively.  At each iteration, we ignore wavelength bins that deviate by $>$3$\sigma$ from the continuum in order to account for unmodeled  absorption and emission features.  We then fit Voigt profiles to the strongest emission lines expected in the spectrum:  \ion{Si}{4}$\lambda$1400, \ion{C}{4}$\lambda$1549, \ion{C}{3}]$\lambda$1909, and \ion{Mg}{2}$\lambda$2799.  Our line wavelengths are taken from the SDSS vacuum wavelength list.\footnote{http://www.sdss.org/dr6/algorithms/linestable.html}  We fit emission lines iteratively as well, ignoring at each step bins that are absorbed by more than $2.5\sigma$ from the continuum $+$ emission fit.

We searched for narrow absorption line (NAL) features attributable to the \ion{Mg}{2}~$\lambda2799$ doublet in each spectrum.  We searched each spectrum from 1550--2799~\AA\ and flagged a candidate \ion{Mg}{2} NAL feature in cases where spectral bins were found at the expected doublet separation which lay at least 4$\sigma$ and 3$\sigma$ below the continuum for the red and blue sides of the doublet, respectively.  In order to screen out broader absorption features, we also required that the spectrum reached a level within $3.5\sigma$ of the continuum in the bin centered between the red and blue components.  We also required that the equivalent width (EW) of at least one of the red and blue components be $<-0.5$~\AA, in order to ensure that the features were reliably detected.  Our results are not strongly dependent on the particular value of the EW threshold.


We searched for \ion{C}{4} BALs by calculating $BI_{0}$, which is similar to the traditional ``balnicity index,'' $BI$, of \citet{wmfh91}, except that it is integrated from velocity offsets --25,000~km~s$^{-1}$ to 0~km~s$^{-1}$ from the QSO rest frame.  $BI$ is roughly a measure of the EW of features that are absorbed across a span of at least 2000~km~s$^{-1}$.  Sources with $BI_0 > 0$ are considered to have BALs.  We tested for \ion{Mg}{2}~$\lambda$~2799 BALs in a similar fashion (when spectra extended to 2800~\AA), although the continuum in that region is difficult to determine due to \ion{Fe}{2} emission and structure in the \ion{Mg}{2} emission line.  We found no cases where \ion{Mg}{2} BALs were present without \ion{C}{4} BALs, and therefore use $BI_0$ measured for the \ion{C}{4} absorption region to distinguish between BAL and non-BAL QSOs.

The UV properties of our sources are listed in Table~\ref{sourceInfoTab} and Table~\ref{emInfoTab}.

\subsection{Fitting X-Ray Spectra\label{fitXRaySec}}

We fit \mbox{X-ray} spectra for each source in order to find and constrain the monochromatic luminosity at 2~keV, $L_{\rm 2~keV}$.  To accomplish this, we fit a broken power law model to each spectrum, with the power law break fixed at 2~keV in the rest frame.  This model accomodates possible soft \mbox{X-ray} absorption or excesses compared to the hard \mbox{X-ray} spectrum.  In cases where the source is not much brighter than the background, we fix $\Gamma = 2$ for the power law fit to the entire \mbox{X-ray} band.  We do this in cases that have fewer than 10 counts in the total (source plus background) spectrum, and also in cases that have no source spectral bins exceeding the background by at least a factor of two when the spectrum is adaptively binned to contain 10 counts per bin.  We fixed $\Gamma = 2$ for 109 sources in our full sample $A$, including 12 of the 139 radio-quiet, non-BAL QSOs in sample $B$ (described in \S\ref{analysisSec}).  We do not use the values of $\Gamma$ in our study, as they can often only be weakly constrained even when they are not fixed to $\Gamma = 2$.

In all cases, we fit the unbinned data using the Cash statistic, $C$ \citep{c79}.  For {\it XMM-Newton} observations, we fit only the spectrum from the camera selected as described in \S\ref{obsCatSec}.  We estimate the 68\% confidence error in $L_{\rm 2~keV}$ by adjusting the model normalization parameter upward and downward, re-fitting the power law photon indices (unless they are fixed to 2 for background-dominated spectra) at each step.  We then estimate the upper and lower limits of $L_{\rm 2~keV}$ from the model fits at the points where $\Delta C = 1$.

Values of $L_{\rm 2~keV}$ for our sources are listed in Table~\ref{sourceInfoTab}.

\subsection{Estimating Radio Loudness\label{estRadLoudSec}}

All but 25 of our 536 sources were covered in the FIRST radio survey \citep{bwh95}.  For these sources, we obtain the 1.4~GHz core flux densities from the DR5 QSO catalog.  For sources that were not covered by the FIRST survey, we used 1.4~GHz flux densities obtained from the NVSS survey \citep{ccgyptb98}.  We estimate the monochromatic luminosity at 5~GHz assuming the radio flux follows a power law with a spectral index $\alpha = -0.8$.  We then calculate the radio-loudness parameter \citep[e.g.,][]{sw80, ksssg89},
\begin{eqnarray}
\log(R^*) &\equiv& \log\biggl( \frac{L_{\nu}({\rm~5~GHz})}{L_{2500\mathring{A}}} \biggr).\label{rStarEqn}
\end{eqnarray}
We classify sources with $\log(R^*) \ge 1$ as ``radio-loud'' and all other sources as ``radio-quiet.''  In 457 of 536 cases, the radio surveys (FIRST or NVSS) were sensitive enough to measure radio luminosities down to $\log(R^*) = 1$.  For 508 cases, the radio data are sensitive down to $\log(R^*) = 1.1$, and in the worst case, the data are only sensitive to $\log(R^*) = 1.5$.  Of the 457 sources with radio observations sensitive down to $\log(R^*) = 1$, 33 are radio-loud (with $\log(R^*) > 1$), and 21 of these have $\log(R^*) > 1.5$.  We therefore estimate that only a small number ($\approx$2) of radio-loud sources have contaminated our ``radio-quiet'' sample, and even these have $\log(R^*) < 1.5$ and therefore are not highly radio-loud.

\section{ANALYSIS OF PHYSICAL PROPERTIES\label{analysisSec}}

We have obtained {\it Chandra} or {\it XMM-Newton} \mbox{X-ray} spectra for 536 SDSS DR5 QSOs at redshifts $1.7 \le z \le 2.7$.  In cases where the source was observed with both {\it Chandra} and {\it XMM-Newton}, we use the {\it Chandra} observation with the longest effective exposure.  We call this set of sources sample $A$.  We use sample $A$ in cases where we are concerned with the general properties of a large number of sources.

Many sources in sample $A$ were not detected in \mbox{X-rays}.  In order to increase the fraction of detected sources, we construct a subsample, which we call sample $B$, of sources observed with {\it Chandra} that are $<$10\arcmin~off-axis and have at least 2.5~ks of exposure.  We do not include {\it XMM-Newton} sources in sample $B$ because the lower sensitivity leads to a larger fraction of \mbox{X-ray} non-detections.  We also limit sample $B$ to contain only sources from the SDSS DR5 QSO catalog that were selected for SDSS spectroscopy based on optical criteria.  These sources include those flagged as ``low-$z$,'' ``high-$z$,'' ``star,'' or ``galaxy'' according to the target selection described in \citet{rfnsvsybbfghikllrsssy02} and \citet{s+02}.  Those sources that were included in the catalog purely due to their radio or \mbox{X-ray} properties are excluded from sample $B$.  Finally, we exclude sources that lie within 1\arcmin~of the {\it Chandra} target in order to remove bias due to observers' tendency to point at unusual sources.  (Dropping this last criterion does not materially affect our results below.)  Sample $B$ contains 163 sources, 139 of which are radio-quiet, non-BAL QSOs.  All sources in sample $B$ were detected by {\it Chandra} in \mbox{X-rays}.  Table~\ref{sampleInfoTab} summarizes the number and types of objects in samples $A$ and $B$.

Sample $B$ is defined without direct reference to source properties and is therefore representative of optically-selected SDSS QSOs.  We use sample $A$ for tests of correlations involving \mbox{X-ray} properties, but use sample $B$ in cases where non-detections adversely affect our statistics.

\subsection{Distribution of $\Delta\alpha_{OX}$\label{dAOXDistSec}}

From our optical/UV and \mbox{X-ray} fits, we calculate the monochromatic luminosities $L_{2500\mathring{A}}$ and $L_{\rm 2~keV}$ at 2500~\AA\ and 2~keV, respectively.  We adopt the traditional definition:
\begin{eqnarray}
\alpha_{OX} &\equiv& 0.3838 \log\biggl( \frac{L_{\rm 2~keV}}{L_{2500\mathring{A}}} \biggr).\label{aoxDefnEqn}
\end{eqnarray}
$\alpha_{OX}$ is a logarithmic measure of the \mbox{X-ray} brightness of a source relative to its UV brightness.  For an ensemble of QSOs, $\alpha_{OX}$ is known to decrease as a function of QSO UV luminosity.  This trend was most recently quantified by \citet{jbssscg07}, who found
\begin{eqnarray}
\alpha_{OX}(L_{2500\mathring{A}}) &=& (-0.140 \pm 0.007) \log(L_{2500\mathring{A}}) + (2.705 \pm 0.212).\label{justDAOXEqn}
\end{eqnarray}

\begin{figure} [ht]
  \begin{center}
      \includegraphics[width=4in, angle=270]{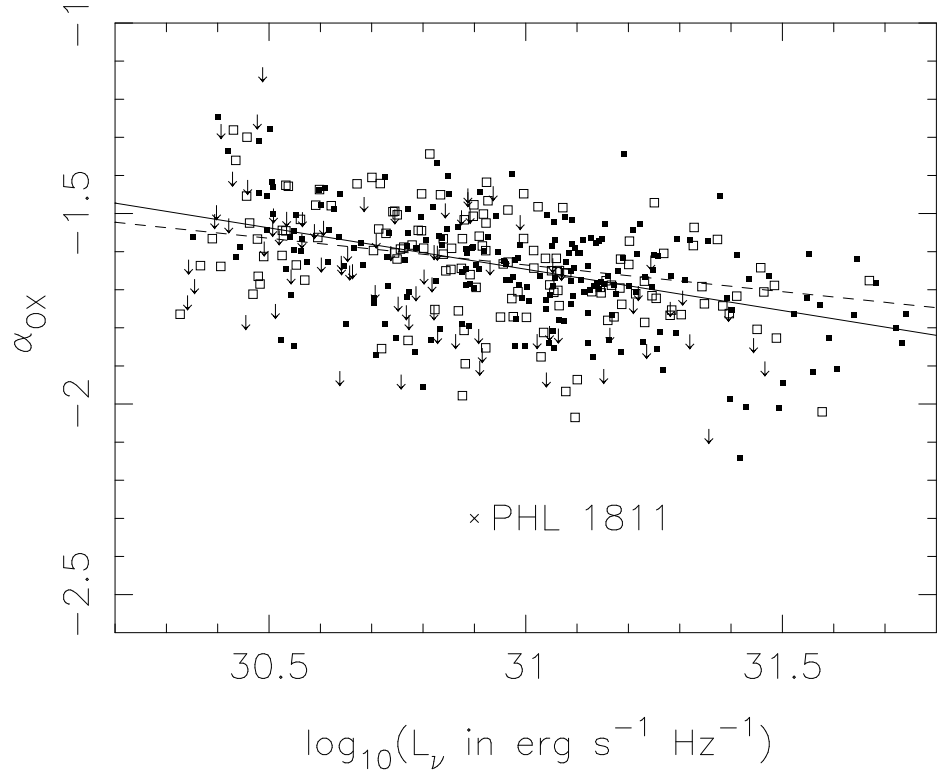}
      \caption{\label{aOXVsLFig}$\alpha_{OX}$ measured for all radio-quiet, non-BAL QSOs plotted against the monochromatic luminosity at 2500~\AA.  Radio-quiet, non-BAL QSOs in the subsample $B$ are all detected and are plotted with open squares, while those detected in $A$ (but not $B$) are plotted with filled squares.  In cases where sources in $A$ were not detected in \mbox{X-rays}, we have plotted upper limits, indicated with arrows.  The dashed line shows the fit that \citet{jbssscg07} obtained from their sample of objects, while the solid line shows the best fit to the radio-quiet, non-BAL QSOs in our sample $B$, described in \S\ref{dAOXDistSec}.  A cross marks the values measured by \citet{lhjgcp07} for the unusual source PHL~1811 (described in \S\ref{introSec}).  PHL~1811 is not included in our SDSS sample, but would be an outlier if it were.}
   \end{center}
\end{figure}

We use the EM (estimate and maximize) regression algorithm \citep{dlr77} in the Astronomy Survival Analysis (ASURV) software package \citep[e.g.,][]{if90, lif92} to fit the trend of $\alpha_{OX}$ with UV luminosity for radio-quiet, non-BAL QSOs in sample $B$.  Since all such sources have \mbox{X-ray} detections in sample $B$, the EM algorithm generates a traditional ``least-squares'' fit \citep{ifn86} in this case.  We find a relation somewhat steeper than Equation~\ref{justDAOXEqn}:
\begin{eqnarray}
\alpha_{OX}(L_{2500\mathring{A}}) &=& (-0.217 \pm 0.036) \log(L_{2500\mathring{A}}) + (5.075 \pm 1.118).\label{myDAOXEqn}
\end{eqnarray}
The values of $\alpha_{OX}$ for our sample, together with the fits from Equations~\ref{justDAOXEqn} and \ref{myDAOXEqn}, are shown in Figure~\ref{aOXVsLFig}.

We then define
\begin{eqnarray}
\Delta\alpha_{OX} &\equiv& \alpha_{OX} - \alpha_{OX}(L_{2500\mathring{A}}),\label{daoxDefnEqn}
\end{eqnarray}
where $\alpha_{OX}$ is the value measured from observations.  $\Delta\alpha_{OX}$ quantifies the relative \mbox{X-ray} brightness of a source compared to that of ``ordinary'' QSOs.  For example, $\Delta\alpha_{OX} = -0.5$ corresponds to \mbox{X-ray} weakness by a factor of $\approx$20 compared to QSOs with the same UV luminosity.

The median of the magnitude of difference in $\alpha_{OX}$ estimated from UV luminosity using Equations~\ref{justDAOXEqn} and \ref{myDAOXEqn} is $0.02$ for our detected sources, which is less than the median measurement error in $\alpha_{OX}$ and is small compared to the range of $\Delta\alpha_{OX}$ in our sample.  The study of \citet{jbssscg07} used a different sample of objects and different methods for estimating luminosities, so the small difference in results is not surprising.  We use Equation~\ref{myDAOXEqn} throughout to determine $\Delta\alpha_{OX}$ in order to correct for any small biases that would cause our results to differ.  We find no evidence of a correlation between $\Delta\alpha_{OX}$ and $z$ for radio-quiet, non-BAL QSOs that might arise from systematic biases in our data.  Because our data cover a relatively small redshift range, this result does not strongly constrain redshift evolution for QSOs in general.

\begin{figure} [ht]
  \begin{center}
      \includegraphics[width=4in, angle=270]{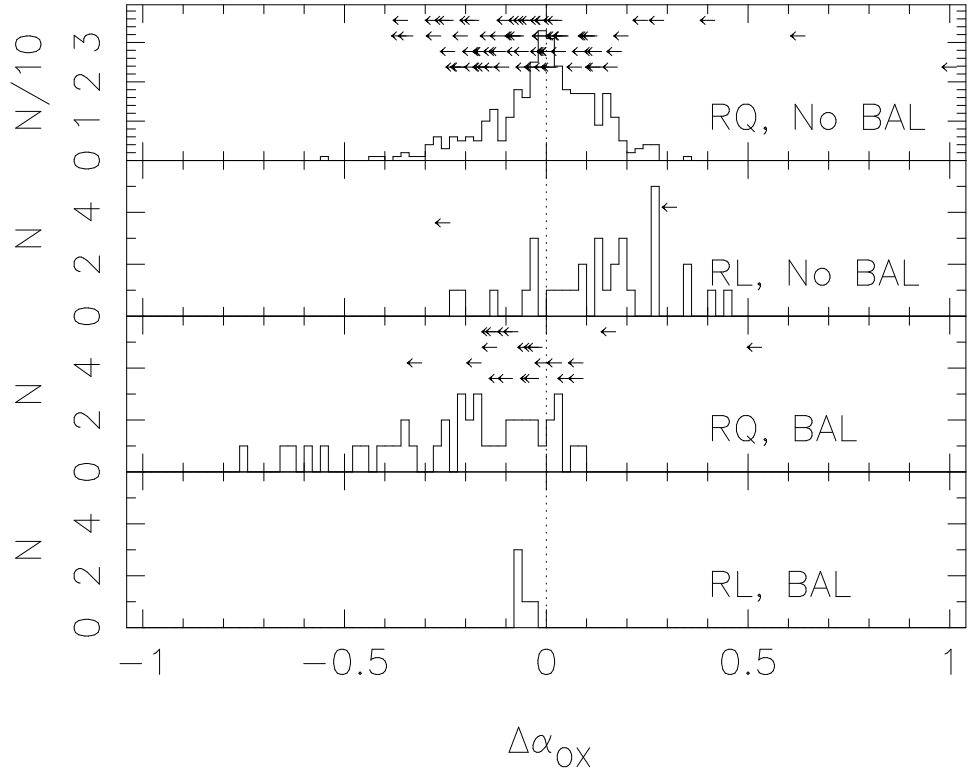}
      \caption{\label{dAOXHistFig}$\Delta\alpha_{OX}$ for the sources in our sample.  From top to bottom, the panels show the distribution of $\Delta\alpha_{OX}$ for radio-quiet sources with no BALs, radio-loud sources with no BALs, radio-quiet sources with BALs, and radio-loud sources with BALs.}
   \end{center}
\end{figure}

\begin{figure} [ht]
  \begin{center}
      \includegraphics[width=4in, angle=270]{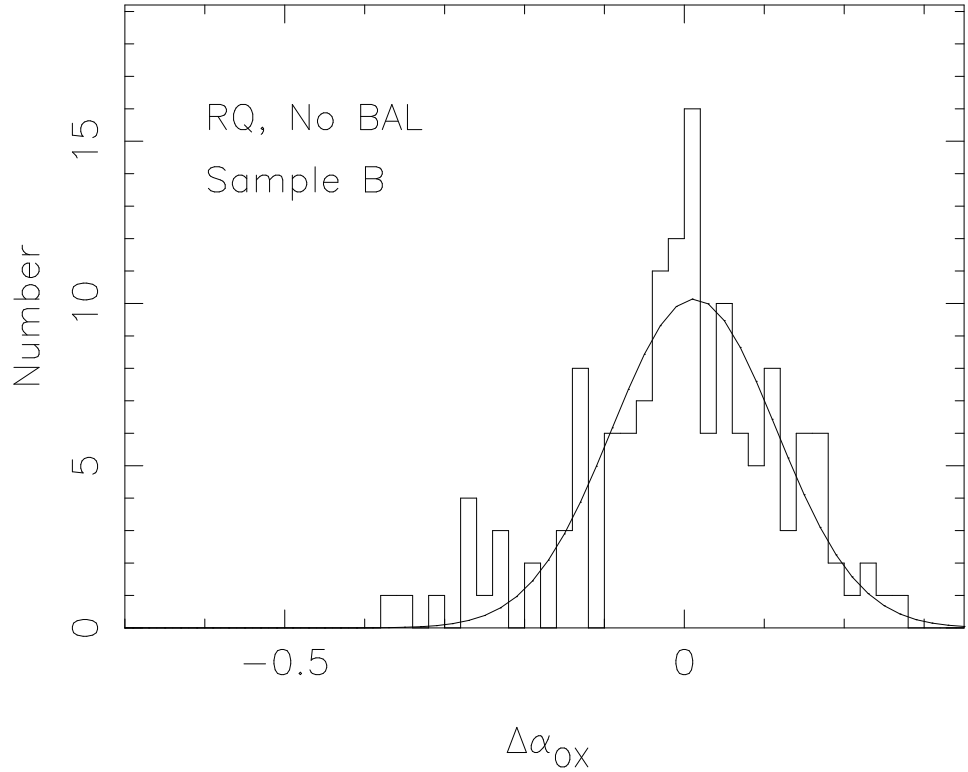}
      \caption{\label{dAOXAEHistFig}$\Delta\alpha_{OX}$ for the radio-quiet sources with no BALs in sample $B$.  The solid line shows the best Gaussian fit to the distribution, discussed in \S\ref{dAOXDistSec}.  The $x$-axis is chosen to match that of Figure~\ref{binomialProbOfltDAOXFig}.}
   \end{center}
\end{figure}

Figure~\ref{dAOXHistFig} shows the distribution of $\Delta\alpha_{OX}$ for our sources.  We divide the sources into four categories, based on the presence (or absence) of a UV BAL and whether they are radio-quiet ($\log(R^*) < 1$) or radio-loud ($\log(R^*) \ge 1$).  Sample $A$ contains 433 radio-quiet, non-BAL QSOs, 34 radio-loud non-BAL QSOs, 64 radio-quiet BAL QSOs, and 5 radio-loud BAL QSOs.  The second panel of Figure~\ref{dAOXHistFig} demonstrates that radio-loud sources are relatively \mbox{X-ray} bright, while the third and fourth panels show that BAL QSOs are \mbox{X-ray} weak compared to non-BAL QSOs.  The distribution of $\Delta\alpha_{OX}$ for radio-quiet, non-BAL QSOs in sample $B$ is shown in Figure~\ref{dAOXAEHistFig}.  An Anderson-Darling test rejects the hypothesis that $\Delta\alpha_{OX}$ is normally distributed for radio-quiet, non-BAL sources in $B$ at 97.5\% confidence.\footnote{The Anderson-Darling test, related to the Kolmogorov-Smirnov test, calculates a statistical probability that the observed data points could come from a normal distribution.  We used case three of the test \citep[e.g.,][]{s74}, in which the mean and width of the (putative) normal distribution are not known {\it a priori}.}

Because the Anderson-Darling test has only indicated non-normality at 97.5\% confidence, we cannot rule out the possibility that the underlying distribution is normal.  In fact, we note that \citet{sbsvv05} found that a Gaussian fit provided a reasonable approximation to the distribution in their larger sample that covered a wider range of UV luminosities and was also slightly contaminated by BAL QSOs.  However, we were not able to test quantitatively the \citet{sbsvv05} distribution for normality due to the \mbox{X-ray} non-detections in their sample.

In order to investigate any possible non-normality of the distribution of $\Delta\alpha_{OX}$ for sample $B$, we have fit the distribution with a Gaussian using the IDL {\tt GAUSSFIT} routine.  The resulting fit, with mean $\mu \approx 0.01$ and $\sigma \approx 0.10$, is shown in Figure~\ref{dAOXAEHistFig}.  Compared to this fit, the measured distribution perhaps has a broader base and a narrow core at $\Delta\alpha_{OX} \approx 0$.  The breadth at the base of the distribution may be due to unmodeled \mbox{X-ray} absorption in non-BAL sources in the \mbox{X-ray} weak wing.  We also speculate that some sources in the \mbox{X-ray} bright wing which we did not consider formally radio-loud may nonetheless have enhanced \mbox{X-ray} emission associated with a modest radio jet.  Detailed study of individual sources is needed to determine the significance of these effects.

\subsection{How Common are Optically-Selected, \mbox{X-ray} Weak QSOs (XWQs)?\label{xWQFreqSec}}

We would like to understand the distribution of relative \mbox{X-ray} brightness/weakness in order to determine the frequency of intrinsically \mbox{X-ray} weak objects.  If strong, intrinsic \mbox{X-ray} emission is not a universal property of QSOs, this would have significant implications for our understanding of AGN physics.  The completeness of \mbox{X-ray} AGN surveys could also be negatively impacted by a significant fraction of intrinsically weak sources.

Because we are able to identify BAL QSOs unambiguously, our sample is well-suited to provide tight constraints on the distribution of relative \mbox{X-ray} brightness/weakness.  All of the 139 optically-selected, radio-quiet, non-BAL sources in sample $B$ are detected in \mbox{X-rays} according to the criteria in \S\ref{obsCatSec}.

The minimum $\Delta\alpha_{OX}$ in sample $B$ is $-0.37$, corresponding to relative \mbox{X-ray} weakness by a factor of $\approx$9.  If the probability of finding a QSO with $\Delta\alpha_{OX} < -0.37$ is $\approx$1.6\%, then there is a $>$90\% chance we would observe at least one such QSOs in 139 attempts, using a binomial distribution.  Since we have observed no such sources, we conclude that $\la$1.6\% of optically-selected, radio-quiet, non-BAL SDSS QSOs have $\Delta\alpha_{OX} < -0.37$.

We have also increased our sample of optically-selected QSOs to include the 87 Palomar-Green (PG) QSOs from the Bright Quasar Survey \citep[BQS;][]{sg83} at $z < 0.5$.  These sources are well-studied in the optical, UV, and \mbox{X-ray} bands.  Figure~\ref{sDSSBQSPlotFig} shows that the BQS sources extend our coverage to lower luminosities as well as lower redshifts.

We have obtained UV and \mbox{X-ray} monochromatic luminosities for BQS sources from Table~2 of \citet{ssbaklsv06}.  We exclude 8 sources known to host BALs.  Five sources ($0043+039$, $1001+054$, $1004+130$, $1700+518$, and $2112+059$) have been identified as BAL QSOs by \citet{blw00} and the authors they reference.  $1351+640$ also hosts a \ion{C}{4} BAL \citep[e.g.,][]{zkwbobdghk01}.  $1126-041$ hosts a variable \ion{C}{4} BAL \citep[][see esp. their Figure~1]{wbwyw99}.  We exclude $1411+442$ \citep[e.g.,][]{mgh87, blw00} and $1535+547$ \citep{ssah97}, both of which have broad absorption at low velocities; the latter is also highly polarized, similar to many BAL QSOs.  For $1259+593$, we use the \mbox{X-ray} monochromatic luminosity obtained with {\it XMM-Newton} measured by \citet{fgrbhv08}; this source was previously undetected in less-sensitive \mbox{X-ray} observations.  It is possible that our remaining sample of BQS QSOs is contaminated with a small number of BAL QSOs.  \citet{blw00} tested for \ion{C}{4} BAL absorption in the spectra 55 of the 87 BQS QSOs at $z < 0.5$.  The X-ray weak QSOs (with $\alpha_{OX} < -1.8$) all had \ion{C}{4} coverage, and in any case BAL QSOs make up a minority \citep[$\approx$15\%; e.g.,][]{hf03} of optically-selected samples, so any any remaining mis-classified BAL QSOs would have only a small influence on our statistics.

There are three BQS sources with $\Delta\alpha_{OX} < -0.4$, and the presence of these sources in our combined sample weakens our constraints on the fraction of \mbox{X-ray} weak QSOs.  The sources are:  PG~$0844+349$ (which is highly \mbox{X-ray} variable), PG~$1011-040$ (which is relatively unabsorbed in \mbox{X-rays}), and PG~$2214+139$ (which is strongly \mbox{X-ray} absorbed).  All three of these sources were discussed in \S\ref{introSec} as examples of unusual \mbox{X-ray} weakness.

With these BQS sources included in our combined sample, we now have 3 sources (out of 217) with $\Delta\alpha_{OX} < -0.4$, corresponding to a constraint that $\la$3.1\% of optically-selected QSOs have $\Delta\alpha_{OX} < -0.4$.  It is not clear why the BQS sources have a greater fraction of \mbox{X-ray} weaker QSOs than we find in our pure SDSS sample.  The cause may be random fluctuations in the data, or it may be due to physical differences in the sources, as the nearby BQS QSOs have lower UV luminosities than the SDSS QSOs in our sample.  In any case, the differences in the upper limits we are able to draw are not great.  The large number of sources in the combined SDSS and BQS sample also allows us to place tight constraints on the weakest sources; we find that $\la$1\% of sources are X-ray weak by a factor of $\ga$56 ($\Delta\alpha_{OX} < -0.67$).

We have given examples of constraints on fractions of sources that are \mbox{X-ray} weaker than some specified value of $\Delta\alpha_{OX}$.  Now we wish to constrain this fraction for arbitrary levels of \mbox{X-ray} weakness.  For the combined sample of 217 SDSS and BQS non-BAL QSOs, we calculated an upper limit on the fraction, $f_{weak}(x)$, of optically-selected QSOs that have $\Delta\alpha_{OX} \le x$.  The distribution of $f_{weak}(x)$ is shown in Figure~\ref{binomialProbOfltDAOXFig} for radio-quiet, non-BAL QSOs in $B$ combined with the BQS QSO sample, and also for the $B$ sample alone.

To generate Figure~\ref{binomialProbOfltDAOXFig}, we used the following method.  For each measured value of $\Delta\alpha_{OX}$ in our sample, we determined the number, $N$, of objects that had a smaller value of $\Delta\alpha_{OX}$.  We used the binomial distribution to find an upper limit on the fraction, $f_{weak}$, of such objects such that we would be 90\% likely to observe at least $N + 1$ objects that are equally \mbox{X-ray} weak or \mbox{X-ray} weaker if the true fraction of such objects is $f_{weak}$.  In Figure~\ref{binomialProbOfltDAOXFig}, we plot the upper limit on the fraction, $f_{weak}(x)$, of objects that have $\Delta\alpha_{OX} \le x$.  

Figure~\ref{binomialProbOfltDAOXByL2500Fig} shows the same analysis performed for the sample of radio-quiet, non-BAL QSOs from $B$ split into three subsamples according to their UV luminosities, $L_{2500\mathring{A}}$.  We draw the tightest constraints from the lowest range of luminosities, although the constraints are highly influenced by differences in the relatively small number of sources with $\Delta\alpha_{OX}\la -0.1$.  We do not find any clear evidence that the fraction of X-ray weak objects depends on luminosity.

\subsection{UV Absorption in Non-BAL QSOs\label{nonBALUVAbsSec}}

BAL QSOs generally have reddened continua compared to those of non-BAL QSOs \citep[e.g.,][]{rrhsvfykb03}.  Apart from the presence of BALs, continuum reddening alone is not known to correlate strongly with \mbox{X-ray} weakness, and \citet{blw00} found no significant difference between the continuum slopes of \mbox{X-ray} weak and non-\mbox{X-ray} weak QSOs.  Using {\it ROSAT} \mbox{X-ray} observations, \citet{sbsvv05} found no trend of redshift-corrected $g-i$ colors with $\alpha_{OX}$.  We find no evidence in our sample for a correlation between $\Delta\alpha_{OX}$ and the ratio between UV flux densities at 1400 and 2500~\AA, which we use as a proxy for intrinsic reddening.  Apart from reddening, this result also suggests that the ratio of \mbox{UV/X-ray} emission is not strongly dependent on the slope of the UV continuum (between 2500 and 1400~\AA) once the trend between UV and X-ray luminosities is factored out with Equation~\ref{myDAOXEqn}.

As described in \S\ref{fitSDSSSec}, we have implemented an algorithm to search for narrow \ion{Mg}{2} absorption doublets that would be associated with intervening systems or with absorbing material intrinsic to the QSO (or its host galaxy).  Figure~\ref{dAOXIntervHistFig} shows the distribution of $\Delta\alpha_{OX}$ for radio-quiet, non-BAL QSOs that have no \ion{Mg}{2} doublets ({\it top panel}) compared to those that have at least one narrow \ion{Mg}{2} doublet ({\it bottom panel}).  We have only considered values for sources at redshift $z < 2.28$ (so that intrinsic \ion{Mg}{2} at rest would be visible in the spectrum) and in which the optical/UV continuum is $>4\sigma$ above zero for at least 50\% of the spectral bins in order to minimize any bias due to the fact that sources with low signal-to-noise may have undetectable absorption lines.  The Gehan and logrank tests implemented in ASURV find no strong evidence that the distributions of $\Delta\alpha_{OX}$ differ between systems with and without narrow \ion{Mg}{2} absorption lines.  If \mbox{X-ray} absorption is related to the presence of narrow \ion{Mg}{2} absorption doublets, the association is more subtle than we can detect with our current sample size, or may be driven by absorption features which are weaker than our EW threshold (\S\ref{fitSDSSSec}).

\subsection{X-Ray Weakness and UV Emission Lines\label{optEmLinesSec}}

In order to investigate any relations between UV emission line properties and \mbox{X-ray} weakness, we test for correlations between $\Delta\alpha_{OX}$ and emission line EW, FWHM, and central wavelength using Kendall's Tau test as implemented in ASURV.  We consider only the \ion{C}{4}~$\lambda 1549$ and \ion{C}{3}]~$\lambda 1909$ emission lines because other lines were not always present in the optical/UV spectra, and also were more difficult to constrain due to nearby spectral structure.  We define the emission line wavelength to be the center of our best-fit Voigt profile.  We omit one source, J$152156.48+520238.4$, from our fits because it has an unusual emission line structure and is an outlier from our sample.  The results of our correlation tests are shown in Table~\ref{lineCorrTestsTab}.  We report the results of correlation tests for radio-quiet sources including and excluding BAL QSOs in the sample.

We see no strong ($>$99\% confidence) correlations between $\Delta\alpha_{OX}$ and the \ion{C}{4}~FWHM, the \ion{C}{3}]~EW, or the \ion{C}{3}] wavelength for radio-quiet sources.  The \ion{C}{4} EW and $\Delta\alpha_{OX}$ are correlated at $>$99.99\% confidence (Figure~\ref{cIVEWVsDAOXFig}).  The \ion{C}{3}] FWHM and $\Delta\alpha_{OX}$ are also significantly anti-correlated at $>99.9$\% confidence (Figure~\ref{cIIIFWHMVsDAOXFig}).  The \ion{C}{4} wavelength is correlated with $\Delta\alpha_{OX}$ at 97--99\% confidence (Figure~\ref{cIVLamVsDAOXFig}), although this correlation weakens if we take the \ion{C}{4} wavelength with respect to the \ion{C}{3}] wavelength rather than 1549.5~\AA\ in the rest frame determined by the SDSS redshift.  Overall, we find that the \mbox{X-ray} weakest sources (including BAL QSOs) have the weakest \ion{C}{4} emission, the broadest \ion{C}{3}] emission lines, and blueshifted \ion{C}{4} emission.

Fitting the trends that were found to have correlations at $\ga$99\% confidence, we obtain:
\begin{eqnarray}
\Delta\alpha_{OX} &=& (0.21 \pm 0.03) \log(CIV~EW) + (-0.36 \pm 0.04),\label{dAOXVsCIVEMEWEqn}
\end{eqnarray}

\begin{eqnarray}
\Delta\alpha_{OX} &=& (-0.15 \pm 0.04) \log(CIII]~FWHM) + (0.19 \pm 0.06), {\rm and}\label{dAOXVsCIIIFWHMEqn}
\end{eqnarray}

\begin{eqnarray}
\Delta\alpha_{OX} &=& (0.004 \pm 0.002) (CIV~\lambda - 1549.5) + (-0.026 \pm 0.009),\label{dAOXVsCIVLamEqn}
\end{eqnarray}
where the EW, FWHM, and central wavelength are in Angstroms.

SDSS redshift determination depends on QSO emission lines, and therefore the redshifts we use may have some dependence on the \ion{C}{4} emission wavelength.  As both line wavelength and $\Delta\alpha_{OX}$ depend on redshift, correlations between the two can be affected by redshift determination.  The degree of change in $\Delta\alpha_{OX}$ due to changes in the redshift should be small compared to the trend in Figure~\ref{cIVEWVsDAOXFig}, but we cannot fully account for the effects of emission line shifts on the SDSS algorithm.  $\Delta\alpha_{OX}$ is also marginally correlated (at $\approx$96\% confidence) with the {\it difference} in the central wavelengths fit to \ion{C}{4} and \ion{C}{3}] emission, and this correlation should be less sensitive to the estimated redshift.  However, because of weaker confidences and redshift determination issues, we do not draw physical conclusions from the observed correlation between emission wavelength and $\Delta\alpha_{OX}$.

In order to illustrate more clearly the effect that correlations with line properties can have on Equation~\ref{myDAOXEqn}, we have repeated the fit described in \S\ref{dAOXDistSec} with the \ion{C}{4}~EW as a second independent variable.  Our best fit is:
\begin{eqnarray}
\alpha_{OX}(L_{\nu},CIV EW) &=& (-0.158 \pm 0.035) \log(L_{2500\mathring{A}}) + (0.210 \pm 0.041) \log(CIV~EW) + (2.925 \pm 1.106).\label{aOXVsLAndCIVEWEqn}
\end{eqnarray}
Equation~\ref{aOXVsLAndCIVEWEqn} characterizes typical \mbox{X-ray} luminosities of optically-selected, radio-quiet, non-BAL QSOs (in the luminosity range of our SDSS sample) based on the UV luminosity and (relatively unknown) physical effects that determine the \ion{C}{4} emission EW.  While the coefficients on $L_{2500\mathring{A}}$ and \ion{C}{4}~EW are both non-zero at $\approx$5$\sigma$ significance, the magnitude of the luminosity term is about fifteen times larger than the EW term for the luminosities and line EWs in our sample.  We also caution that the UV luminosity range in our sample is limited, and we cannot be sure that Equation~\ref{aOXVsLAndCIVEWEqn} extends to other UV luminosities.

\subsection{Constraints on UV/\mbox{X-ray} Variation\label{uVXVarConstraintSec}}

Even when radio-loud and BAL QSOs are removed from samples, there is still a significant amount of scatter about the best-fit UV/\mbox{X-ray} luminosity relation (Equation~\ref{myDAOXEqn}).  The scatter may be caused by additional physical factors, some of which have been considered in \S\ref{nonBALUVAbsSec} and \S\ref{optEmLinesSec}.  In this section, we place a limit on the maximal extent to which the UV and \mbox{X-ray} luminosities may vary with respect to each other.  To determine this limit, we make the ``worst-case'' assumption that the scatter is entirely caused by source variability.  We also assume for statistical purposes that $\Delta\alpha_{OX}$ is normally distributed, even though an Anderson-Darling test suggests this may not be perfectly accurate (\S\ref{dAOXDistSec}).  Figure~\ref{dAOXAEHistFig} shows that the true distribution does not differ too strongly from a normal distribution, and this assumption should not greatly affect our result.  Finally, our limit is also weakened by the non-simultaneity of our UV and X-ray measurements.  Even if the UV and X-ray luminosities vary synchronously, non-simultaneous measurements will show increased scatter over simultaneous observations.  For all these reasons, our analysis only gives an upper limit on the relative variation between UV and X-ray luminosities.

Following \citet{mgwzs88}, we have minimized the likelihood function
\begin{eqnarray}
L &\equiv& \prod \frac{1}{\sqrt{2\pi(\sigma^2_i + \sigma^2)}} \exp(-(\Delta\alpha_{OX,i} - \mu)^2 / 2(\sigma^2_i + \sigma^2)))\label{maxLikDAOXEqn}
\end{eqnarray}
in order to determine the intrinsic mean ($\mu$) and rms ($\sigma$) for our $\Delta\alpha_{OX}$ values, each having an estimated measurement error $\sigma_i$.  For radio-quiet, non-BAL QSOs in sample $B$, we find $\mu=0.005 \pm 0.015$ and $\sigma = 0.115 \pm 0.011$.  In fact, the square root of the unbiased sample variance is $0.125$, demonstrating that the influence of the measurement error is small compared to the spread of $\Delta\alpha_{OX}$.  The careful (unbiased) construction of sample $B$ has allowed us to improve constraints on the spread of $\Delta\alpha_{OX}$; our $\sigma$ is significantly smaller than the $\alpha_{OX}$ rms of $\approx$0.2 found by \citet{ssbaklsv06} for QSOs in their sample with $30 < \log(L_{2500\mathring{A}}) < 31$.

Assuming the width of the distribution of $\Delta\alpha_{OX}$ is purely due to luminosity variation (and measurement error), we can use the intrinsic rms to constrain QSO \mbox{X-ray} fluxes to vary upward or downward with respect to UV fluxes by a factor of only about $10^{0.115 / 0.3838} \approx 2$ for sources within $1\sigma$ of the mean, $\mu$.  For sources within 2$\sigma$ and 3$\sigma$ of the mean, this corresponds to factors of $\approx$4 and $\approx$8, respectively.  This estimate only accounts for \mbox{X-ray} variation with respect to the UV continuum.  UV fluxes may also be varying along with the \mbox{X-ray} fluxes, so that the absolute \mbox{X-ray} variation is larger.  Compared to this upper limit, \mbox{X-ray} variation by a factor of 30, such as that seen in Mrk~335 \citep{gkg07}, is rather unusual, unless the UV continuum has also varied strongly.  However, we note that our SDSS sample does not extend to the lower UV luminosity \citep[$L_{2500~\mathring{A}} \sim 10^{29}$~erg~s$^{-1}$~Hz$^{-1}$,][]{gkg07} of this source, and that variability properties may differ at lower luminosities.

We have also performed this analysis calculating $\Delta\alpha_{OX}$ using Equation~\ref{aOXVsLAndCIVEWEqn} in order to remove scatter in the UV/\mbox{X-ray} luminosity relation that is correlated with the \ion{C}{4} emission EW.  These effects are small, however, compared to the overall span of UV and \mbox{X-ray} luminosities, and the end result is only a slightly tighter constraint on variation.  We obtain in this case $\mu=0.009 \pm 0.014$ and $\sigma = 0.103 \pm 0.011$.

Nearby Seyfert 1 AGN typically show fractional variability in their X-ray luminosities by up to $\approx$40\% in \mbox{X-rays} over a few years \citep{mev03}.  Apart from this general study, dramatic variation has been observed in individual sources such as NGC~4051 \citep[e.g.,][]{lmuj03}.  The variability characteristics of higher-luminosity AGN are not well-understood; recent studies have found some evidence that QSOs vary less at higher luminosities, but that variability increases at $z \approx 2$ \citep[e.g.,][]{mal02, psgkg04}.  In individual cases, variation can be extreme. \citet{sbvsfrs05} observed luminosity variation by factors of $\approx$2--4 over 3--5 (rest frame) months in several QSOs at $z > 4$.

In order to test whether variability could account for the spread of $\Delta\alpha_{OX}$, we assume a UV variability amplitude of $\approx$30\% \citep{gmkns99, kbmnss07} that is uncorrelated to the \mbox{X-ray} variation ($\sim$40\%).  We caution that these are simply estimates and that the variation properties are poorly constrained.  In some individual cases, high-luminosity QSOs have shown variation of up to 70\% in the UV over 2--3~yr in the rest-frame \citep{kbmnss07}, and in X-rays by a factor of $\approx$4 over 73~days in the rest-frame \citep{sbvsfrs05}.  Using these estimates, the width of the distribution of $\Delta\alpha_{OX}$ is largely (70--100\%) attributable to intrinsic variation.

\section{DISCUSSION\label{discussionSec}}

In this section, we discuss additional implications of the results presented in \S\ref{analysisSec}, including:  the rarity of extremely \mbox{X-ray} faint, optically-selected QSOs (\S\ref{xSDSSDiscSec}); the relation of the anomalously \mbox{X-ray} faint QSO PHL~1811 to our SDSS sample (\S\ref{pHL1811Sec}); and the dependence of UV emission line strengths on X-ray luminosity (\S\ref{optEmLineDiscSec}).

\subsection{X-Ray Weakness in Optically-Selected QSOs\label{xSDSSDiscSec}}

Using \mbox{X-ray} data from the {\it Einstein Observatory}, \citet{at86} derived an upper limit of 8\% (at 95\% confidence) for the fraction of optically-selected QSOs that could be so \mbox{X-ray} faint as to be a separate population from ``ordinary,'' \mbox{X-ray}-emitting QSOs.  At the time of their study, it was not known that BAL QSOs were anomalously X-ray weak.  However, Figure~\ref{dAOXHistFig} demonstrates that BAL QSOs generally live on the tail of the overall distribution of relative X-ray brightness, and would likely not be considered a disjoint population by the criteria of \citet{at86}.

The high sensitivity of {\it Chandra} has enabled us to constrain the fraction of QSOs that lie in the \mbox{X-ray}-weak tail of the distribution of ``ordinary'' QSOs.  If a separate population of extremely \mbox{X-ray} weak QSOs exists, as in the hypothesis tested by \citet{at86}, our results also provide upper limits on the fraction of such sources in optically-selected samples.  Although we cannot rule out the possibility that a small fraction of QSOs are extremely intrinsically \mbox{X-ray} faint, we have determined that such sources are at least very unusual in the SDSS.

In fact, we find evidence that few ($\la$2\%) SDSS QSOs are relatively \mbox{X-ray} weak by even a factor of 10.  When BQS QSOs are included in the sample, we find that $\la$1\% of sources are X-ray weak by a factor of $\la$56 ($\Delta\alpha_{OX} < -0.67$).  Thus, if accretion disk coronae are responsible for \mbox{X-ray} emission in QSOs, they appear to be present and effective in most, if not all, luminous AGN.  Figure~\ref{binomialProbOfltDAOXFig} shows our estimates of upper limits on the fraction of radio-quiet, non-BAL QSOs that are \mbox{X-ray} weaker than a given value of $\Delta\alpha_{OX}$.  These limits are driven by the sizes of our samples.

Figure~\ref{binomialProbOfltDAOXFig} allows the calibration of wide, shallow \mbox{X-ray} surveys to constrain the fraction of optically-selected QSOs that should be detected.  The current upper limits indicate that at most a few percent of optically-selected radio-quiet, non-BAL QSOs will be missed at $\Delta\alpha_{OX} < -0.4$.  We have achieved a 100\% detection fraction for the sources in sample $B$ with exposure times as short as 2.5~ks.  Future shallow, wide \mbox{X-ray} surveys should therefore be able to detect nearly all non-BAL QSOs with UV luminosities as faint as those in the SDSS in relatively short exposure times.

\subsection{How Unusual Is PHL~1811?\label{pHL1811Sec}}

A multiwavelength study of the nearby ($z=0.192$) narrow-line QSO PHL~1811 has revealed that this source has both unusual \mbox{X-ray} and optical/UV spectra \citep{lhjgcp07, lhjc07}.  It has $\Delta\alpha_{OX} \approx -0.7$, corresponding to \mbox{X-ray} weakness by a factor of almost 70 compared to QSOs with the same UV luminosity.  There is little, if any, intrinsic \mbox{X-ray} absorption, the UV spectrum shows no broad absorption lines, and broad optical/UV emission lines are very weak.  \ion{C}{4}~$\lambda$1549 has an EW of only 6.6~\AA, and only a limit of $<$1.7~\AA\ can be placed on the \ion{C}{3}] EW.

If PHL~1811 were at a redshift $z = 2$, its continuum properties would justify its inclusion in our sample.  Its blue color would identify it as a QSO candidate for SDSS spectroscopy, according to the criteria specified by \citet{rfnsvsybbfghikllrsssy02}, and \citet{lhjc07} note (in their \S4.5) that it would have been identified as a PG QSO, if it had been in that survey's footprint.  The 2500~\AA\ monochromatic luminosity of PHL~1811 is $L_{2500\mathring{A}} \approx 8\times 10^{30}$~erg~s$^{-1}$~Hz$^{-1}$ \citep[][\S4]{lhjgcp07}, which is intermediate for our sources (Figure~\ref{aOXVsLFig}).  Thus, our survey can be used to constrain the frequency of intrinsically \mbox{X-ray} weak objects like PHL~1811 in the Universe.  From Figure~\ref{binomialProbOfltDAOXFig}, it appears that the weak intrinsic \mbox{X-ray} emission of PHL~1811 makes it an unusual source that is not representative of a significant fraction of optically-selected QSOs.  If the X-ray weakness of PHL~1811 is due to an abnormal accretion disk corona, as discussed in \S5.1 of \citet{lhjgcp07}, such physical states are apparently rare in optically-selected AGN.

The weak, blueshifted \ion{C}{4}~$\lambda$1549 emission observed in PHL~1811 is qualitatively consistent with the trends we observe for \mbox{X-ray} weaker objects (\S\ref{optEmLinesSec}).  However, it is still an outlier compared to the emission line properties of our sample.  For example, Equation~\ref{cIVEMEWVsLUVXEqn} predicts a \ion{C}{4} emission EW of 17~\AA\ based on UV and \mbox{X-ray} luminosities, while \citet{lhjc07} actually measure 6.6~\AA.  If the \mbox{X-ray} luminosity were to increase, this would also increase the predicted emission line strength, making the discrepancy worse.  PHL~1811 is therefore an outlier from our sample according to both continuum and emission-line properties, for physical reasons that are currently unknown.

\subsection{Broad Emission Lines Correlated with X-Ray Brightness\label{optEmLineDiscSec}}

While the strong UV absorption in BAL QSOs is known to be associated with \mbox{X-ray} weakness, it is not clear how strongly other properties evident in the optical/UV spectrum are associated with \mbox{X-ray} weakness.  We do not find any strong correlations between \mbox{X-ray} weakness and UV reddening or the presence of narrow absorption line systems.  We do find significant correlations between $\Delta\alpha_{OX}$ and the \ion{C}{4} emission EW, the \ion{C}{3}] emission FWHM, and the wavelength shift of the \ion{C}{4} emission line.

It has been known for some time that \ion{C}{4} emission EWs are anti-correlated with the UV continuum strength.  This is the well-known Baldwin effect \citep{b77}, typically parameterized as $EW \propto L_{UV}^{\beta}$, where $L_{UV}$ is the UV luminosity.  For an ensemble of QSOs, the ``global Baldwin effect'' has a logarithmic slope $\beta \approx -0.2$ \citep{krk90}.  For a sample of 105 SDSS QSOs, \citet{wvbb06} find $\beta = -0.22 \pm 0.03$.  For our sample, we find $\beta = -0.24 \pm 0.03$, a bit steeper than but consistent with previous studies.  The comparison is complicated by the fact that \citet{krk90} measured the UV luminosity at 1550~\AA\ rather than at 2500~\AA, but comparison of these luminosities for our data indicates that the parameter $\beta$ should not be greatly affected.

Equation~\ref{dAOXVsCIVEMEWEqn} indicates that the \ion{C}{4} emission EW may also be related to \mbox{X-ray} luminosities.  This is not unexpected in a photionization-driven scenario, as the K and L shells of \ion{C}{4} have ionization potentials of $\approx$320 and $\approx$64~eV, respectively.  Fitting the emission EW as a function of both UV and \mbox{X-ray} luminosity for sample $B$, we find:
\begin{eqnarray}
\log(CIV EW) &=& (-0.41 \pm 0.07)\log(L_{2500\mathring{A}}) + (0.29 \pm 0.06)\log(L_{\rm 2~keV}) + (6.38 \pm 2.09)\label{cIVEMEWVsLUVXEqn}.
\end{eqnarray}
According to Equation~\ref{cIVEMEWVsLUVXEqn}, when \mbox{X-ray} luminosities are considered, the dependence on UV luminosities steepens somewhat to $\beta_{UV} \approx -0.41$.  The coefficient on the \mbox{X-ray} luminosity is also significant, with $\beta_X \approx 0.29$.  The two coefficients have opposite signs, so that \ion{C}{4} emission weakens with increasing UV luminosity, but increases with \mbox{X-ray} luminosity.

By definition, $\alpha_{OX}$ increases as $L_{\rm 2~keV}$ increases or $L_{2500\mathring{A}}$ decreases.  The opposite signs and roughly comparable magnitudes of the coefficients $\beta_{X}$ and $\beta_{UV}$ in Equation~\ref{cIVEMEWVsLUVXEqn} indicate that the \ion{C}{4} emission strength is correlated with $\alpha_{OX}$.  In fact, \citet{g98} previously observed that the \ion{C}{4} emission EW depended more strongly on $\alpha_{OX}$ than on UV luminosity alone, and suggested that the emission line strength may therefore be driven by the continuum shape (e.g., the slope between UV and X-ray luminosities characterized by $\alpha_{OX}$).  A correlation between \ion{C}{4} EW and the ratio of monochromatic luminosities at 1~keV and 1350~\AA\ has also been found by \citet{wlz98}.  Such observations correspond well with photoionziation models of equilibrated gas, in which the shape of the ionizing continuum is an important parameter for determining ionization fractions.  In fact, it has been argued that the Baldwin effect may be driven by the photoionizing flux \citep[e.g.,][]{kg04}, although further modeling is needed to determine whether the X-ray dependence of the Baldwin effect is consistent with photoionization scenarios.

On the other hand, \ion{C}{4} emission line properties are known to be related to the Eddington ratio, $L / L_{Edd}$ \citep[e.g.,][]{bl05}.  The $H_{\beta}$ line is redshifted out of the SDSS spectral range, so we are unable to determine Eddington ratios for our sources using $H_{\beta}$ line diagnostics.  As an alternative to photoionzation scenarios, the relation between \ion{C}{4}~EW and \mbox{UV/X-ray} luminosities may be due to an underlying physical parameter, such as Eddington ratio or perhaps orientation \citep[e.g.,][]{rvrhssty02, grhbsv05}, that drives the \ion{C}{4} emission and both the UV and \mbox{X-ray} luminosities.  It has also been suggested that \ion{C}{4} blueshifts, the Baldwin effect, and relative \mbox{X-ray} brightness have a common physical origin, and that the underlying physical parameter governs a spectrum of QSO properties.  BAL QSOs would fall on the weak-emission extreme of this spectrum \citep[][and references therein]{r06}.

\section{CONCLUSIONS AND FUTURE WORK\label{concSec}}

\begin{figure} [ht]
  \begin{center}
      \includegraphics[width=4in, angle=270]{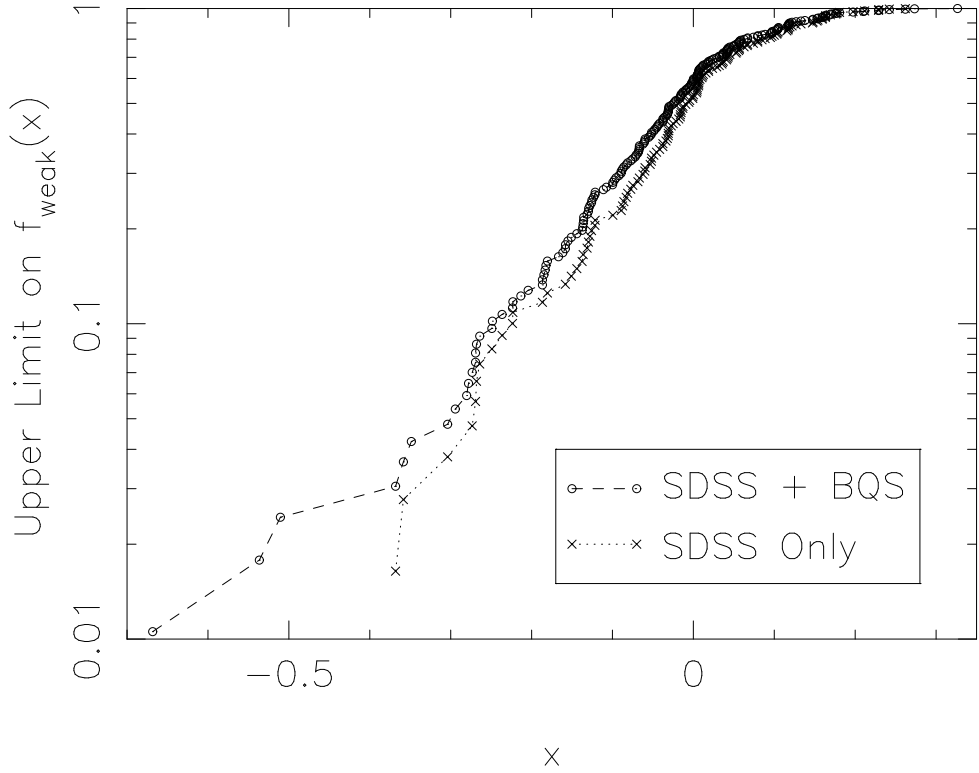}
      \caption{\label{binomialProbOfltDAOXFig}Upper limits on the fraction $f_{weak}(x)$ of sources which have $\Delta\alpha_{OX} \le x$, calculated as described in \S\ref{xWQFreqSec}.  The dashed line connecting circles shows results for sample $B$ combined with BQS QSOs.  The dotted line connecting crosses shows results for sample $B$ alone.  The $y$-axis is logarithmic.}
   \end{center}
\end{figure}

\begin{figure} [ht]
  \begin{center}
      \includegraphics[width=4in, angle=270]{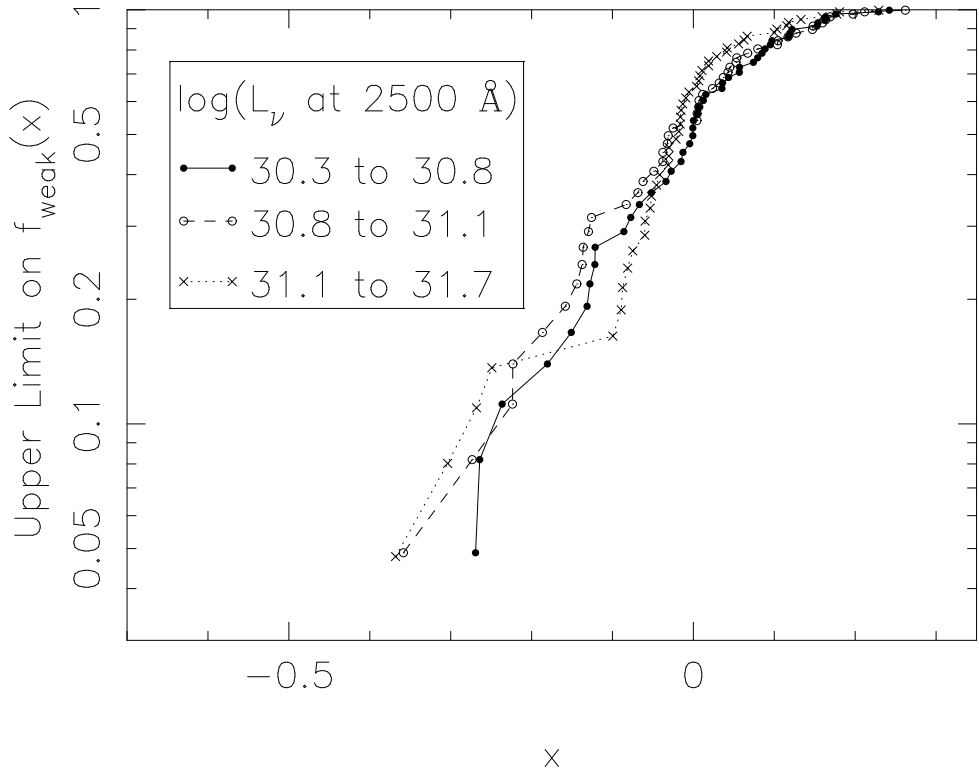}
      \caption{\label{binomialProbOfltDAOXByL2500Fig}Similar to Figure~\ref{binomialProbOfltDAOXFig}, we plot upper limits on the fraction $f_{weak}(x)$ of sources which have $\Delta\alpha_{OX} \le x$.  We use radio-quiet, non-BAL QSOs from sample $B$, and split the combined sample into three subsamples according to $L_{2500\mathring{A}}$.  The key shows the plot symbols that correspond to each UV luminosity range.  The $y$-axis is logarithmic.}
   \end{center}
\end{figure}

\begin{figure} [ht]
  \begin{center}
      \includegraphics[width=4in, angle=270]{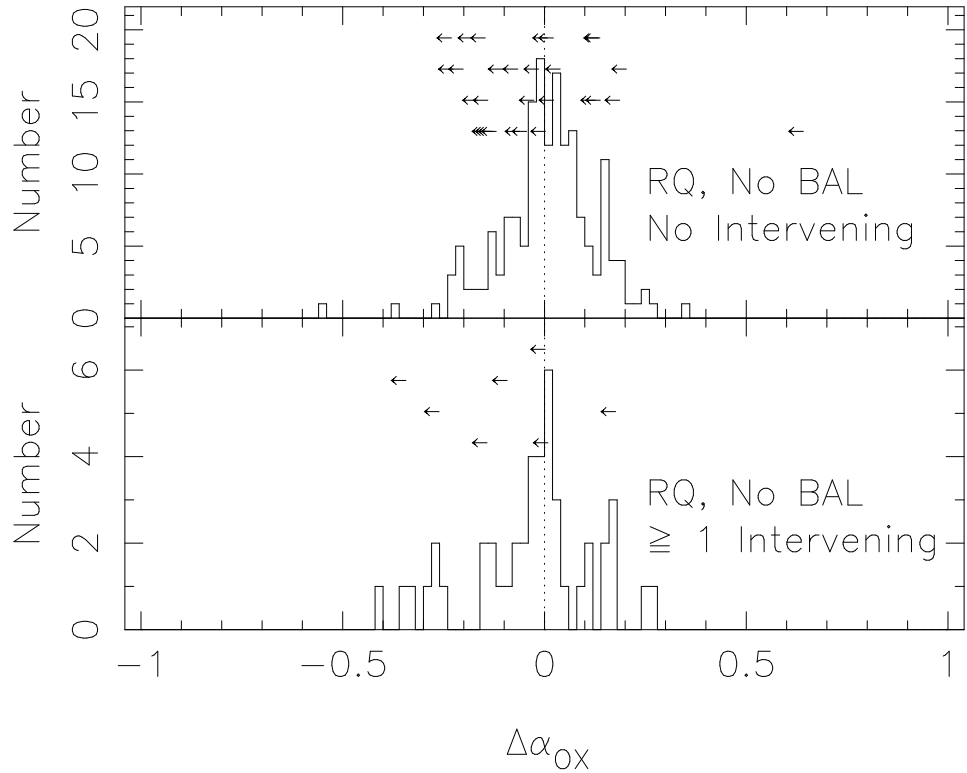}
      \caption{\label{dAOXIntervHistFig}The distribution of $\Delta\alpha_{OX}$ for radio-quiet QSOs with no BAL and no identified narrow-line \ion{Mg}{2} absorbers ({\it top panel}) and for sources with at least one narrow-line absorber ({\it bottom panel}).}
   \end{center}
\end{figure}

\begin{figure} [ht]
  \begin{center}
      \includegraphics[width=4in, angle=270]{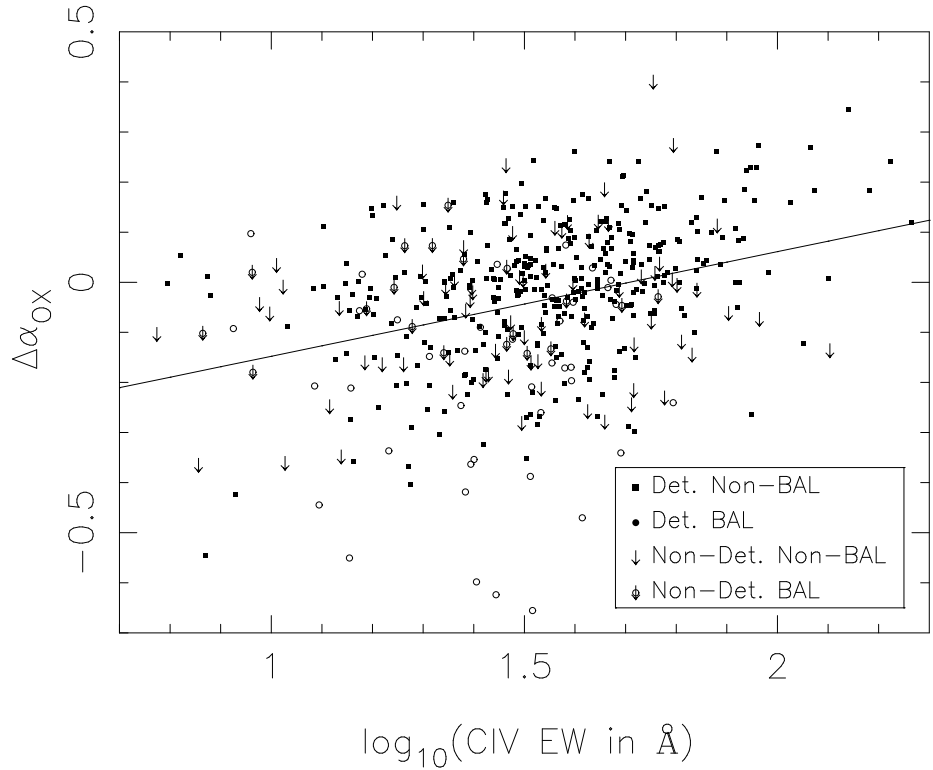}
      \caption{\label{cIVEWVsDAOXFig}$\Delta\alpha_{OX}$ plotted against rest-frame EWs for \ion{C}{4} emission lines.  All detected, non-BAL, radio-quiet QSOs are plotted as filled squares, with upper limits given by arrows.  Detected, radio-quiet BAL QSOs are plotted as open circles, with upper limits for non-detected radio-quiet BAL QSOs designated by arrows inside the circles.  The plot ranges clip off a small number of points.  The solid line indicates the best fit to the data.}
   \end{center}
\end{figure}

\begin{figure} [ht]
  \begin{center}
      \includegraphics[width=4in, angle=270]{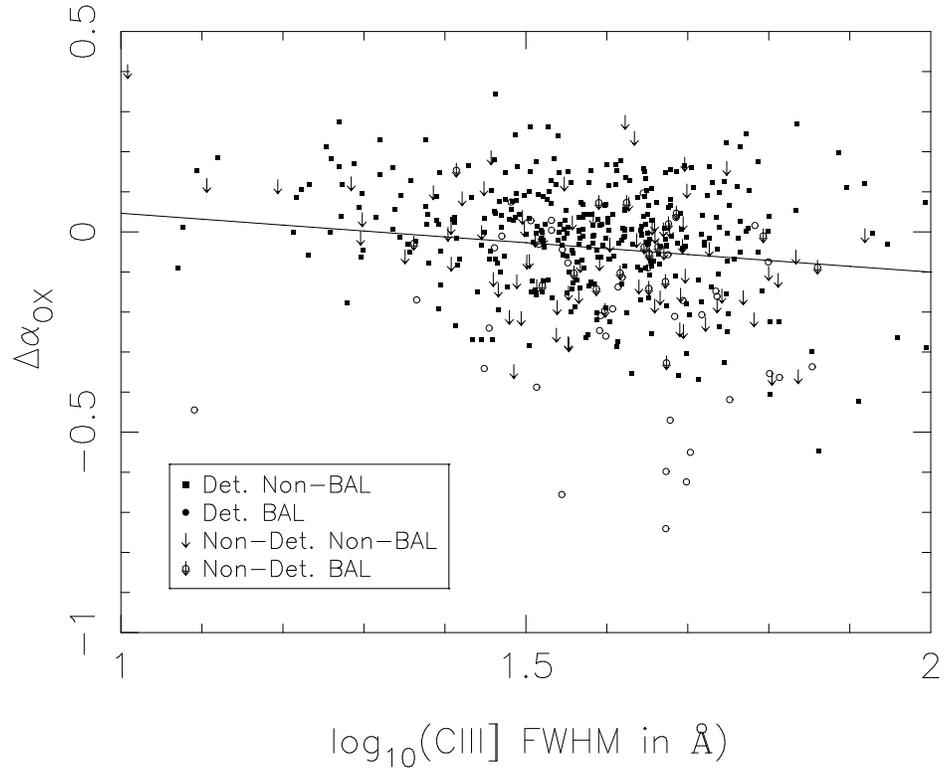}
      \caption{\label{cIIIFWHMVsDAOXFig}Same as Figure~\ref{cIVEWVsDAOXFig}, but showing the FWHM of \ion{C}{3}]~$\lambda$1908 on the $x$-axis.}
   \end{center}
\end{figure}

\begin{figure} [ht]
  \begin{center}
      \includegraphics[width=4in, angle=270]{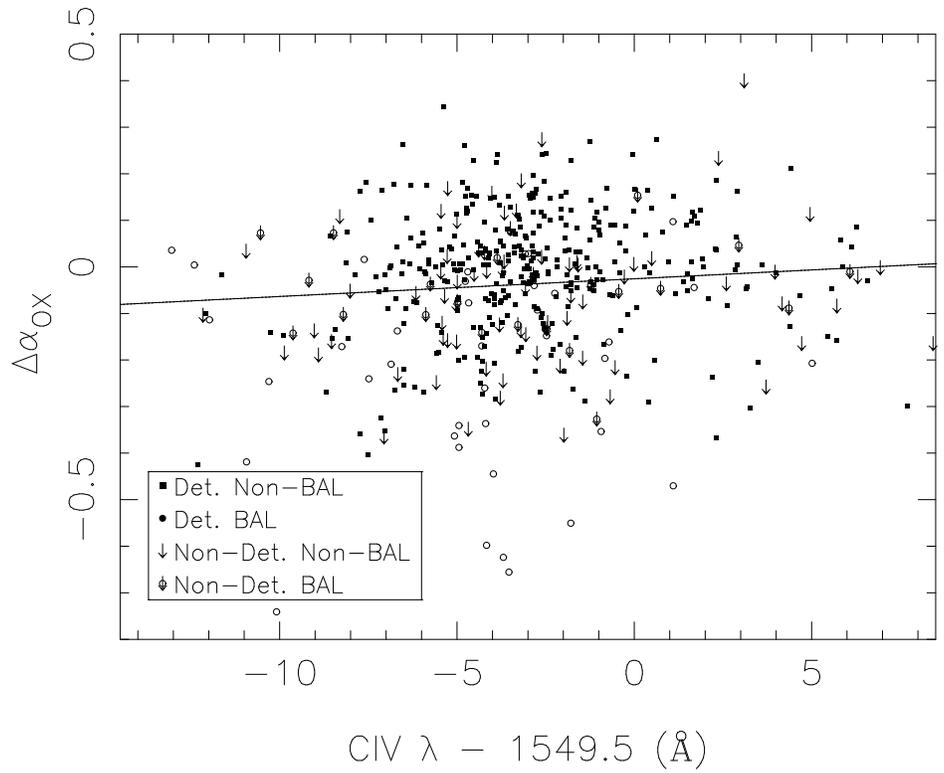}
      \caption{\label{cIVLamVsDAOXFig}Same as Figure~\ref{cIVEWVsDAOXFig}, but showing the central wavelength of \ion{C}{4}~$\lambda$1549 on the $x$-axis.}
   \end{center}
\end{figure}

We have analyzed the relationship between \mbox{X-ray} luminosity and UV properties of SDSS DR5 QSOs at redshift $1.7 \le z \le 2.7$.  At these redshifts, we are able to identify \ion{C}{4} BAL QSOs in our sample.  This has enabled us to place strong constraints on the fraction of intrinsically \mbox{X-ray} weak QSOs, as well as to search for additional trends beyond the well-studied relationship between UV and \mbox{X-ray} luminosities.  In particular, we find that:
\begin{enumerate}
\item{Non-BAL, radio-quiet QSOs which are very (intrinsically) \mbox{X-ray} weak are rare in optically-selected samples, with $\la$2\% of SDSS QSOs having $\Delta\alpha_{OX} < -0.4$ and $\la$1\% of SDSS/BQS QSOs having $\Delta\alpha_{OX} < -0.67$.  Figure~\ref{binomialProbOfltDAOXFig} provides upper limits on the fraction of sources that may be relatively \mbox{X-ray} weaker than a given value of $\Delta\alpha_{OX}$.}
\item{The rms of the $\Delta\alpha_{OX}$ distribution for radio-quiet, non-BAL QSOs is about $0.1$, corresponding to a factor of $\approx$2 spread in \mbox{X-rays} relative to that estimated from the UV luminosity.  This places an upper limit on typical \mbox{X-ray} variability with respect to the UV continuum.}
\item{While the amplitude of and relation between UV and X-ray variation in QSOs is not well-understood, estimates based on recent studies indicate that most, if not all, of the observed spread of $\Delta\alpha_{OX}$ can be attributed to variability.}
\item{The distribution of $\Delta\alpha_{OX}$ may not be normally distributed (at 97.5\% confidence) in our sample $B$.  Perhaps physical processes such as \mbox{X-ray} absorption are significant in some sources, broadening the wings of the distribution.}
\item{We find no strong evidence that reddening or the presence of narrow \ion{Mg}{2} absorption is related to \mbox{X-ray} weakness for non-BAL QSOs.}
\item{UV emission line properties such as \ion{C}{4}~EW, \ion{C}{3}]~FWHM, and perhaps \ion{C}{4} wavelength are correlated to relative \mbox{X-ray} brightness.}
\item{The \ion{C}{4} emission EW depends on both UV and \mbox{X-ray} luminosity.  The physics that drives the global Baldwin effect is apparently associated with \mbox{X-ray} emission as well as UV emission.}
\item{Even after correcting for secondary trends (such as weak \ion{C}{4} emission, which is associated with relative \mbox{X-ray} weakness), objects that are as intrinsically \mbox{X-ray} faint as PHL~1811 are rare.}
\end{enumerate}

We have quantitatively shown that luminous X-ray emission is essentially a universal property of optically-selected QSOs.  Future studies can test whether this result holds true for QSOs selected in other wavebands, such as the radio and infrared.  New optical surveys will extend the luminosity range of our sample, while UV (and infrared) spectroscopy will allow BAL QSOs to be identified (and $L_{2500~\mathring{A}}$ to be measured) for sources in a wider redshift range.  The high detection rate for relatively short exposures in our sample has also demonstrated that wide, shallow X-ray surveys at high angular resolution are an effective way to study the X-ray properties of bright, optically-selected QSOs.

\acknowledgements
We gratefully acknowledge support from NASA LTSA grant NAG5-13035.  Most of the data analysis for this project was performed using the ISIS platform \citep{hd00}.  We thank the referee for helpful comments.

Funding for the SDSS and SDSS-II has been provided by the Alfred P. Sloan Foundation, the Participating Institutions, the National Science Foundation, the U.S. Department of Energy, the National Aeronautics and Space Administration, the Japanese Monbukagakusho, the Max Planck Society, and the Higher Education Funding Council for England.  The SDSS Web site \hbox{is {\tt http://www.sdss.org/}.}

Most of the data analysis for this project was performed using the ISIS platform \citep{hd00}.  We thank Ohad Shemmer for helpful discussions of QSO variability.


\bibliographystyle{apj3}
\bibliography{apj-jour,bibliography}

\clearpage

\begin{landscape}
\begin{deluxetable}{lrrrrrrrrrrrrrr}
\tabletypesize{\scriptsize}
\tablecolumns{12}
\tablewidth{0pc}
\tablecaption{Source Information\label{sourceInfoTab}\tablenotemark{a}}
\tablehead{\colhead{SDSS} & \colhead{$z$} & \colhead{ObsId\tablenotemark{b}} & \colhead{X-Ray} & \colhead{X-Ray Counts} & \colhead{X-Ray} & \colhead{Sample} & \colhead{$\log(R^*)$} & \colhead{C\sc{IV} $BI_0$} & \colhead{Narrow Mg\sc{II}} & \colhead{$\log(L_{1400\mathring{A}})$\tablenotemark{h}} & \colhead{$\log(L_{2500\mathring{A}})$\tablenotemark{h}} & \colhead{$\log(L_{\rm 2~keV})$\tablenotemark{h}} & \colhead{$\alpha_{OX}$\tablenotemark{i}} & \colhead{$\Delta\alpha_{OX}$} \\ \colhead{Source} & \colhead{} & \colhead{} & \colhead{Exposure (ks)\tablenotemark{c}} & \colhead{Soft/Hard\tablenotemark{d}} & \colhead{Detected?\tablenotemark{e}} & \colhead{$B$?\tablenotemark{f}} & \colhead{} & \colhead{(km s$^{-1}$)} & \colhead{Systems\tablenotemark{g}} & \colhead{} & \colhead{} & \colhead{} & \colhead{}}
\startdata
J$000654.10-001533.4$ & 1.73 & 4096 & 4.5 & 34/10 & 1 & 0 & $<0.19$ &     0 & 3 & 31.01 & 31.23 & 26.89 & $-1.67$(0.03) & $0.03$\\
J$000659.28-001740.8$ & 2.02 & 4096 & 4.5 & 19/5 & 1 & 1 & $<0.75$ &     0 & 0 & 30.70 & 30.80 & 26.77 & $-1.55$(0.04) & $0.06$\\
J$000717.73-002811.3$ & 1.89 & 4096 & 4.5 & 11/0 & 1 & 0 & $2.18$ &     0 & 0 & 30.03 & 30.31 & 27.12 & $-1.22$(0.07) & $0.27$\\
J$001130.55+005550.7$ & 2.31 & 0403760301 & 25.4 & 34/29 & 1 & 0 & $<0.28$ &   928 & 0 & 31.02 & 31.37 & 26.33 & $-1.94$(0.03) & $-0.21$\\
J$001247.12+001239.5$ & 2.15 & 4829 & 6.7 & 9/3 & 1 & 1 & $<0.54$ &     0 & 0 & 30.82 & 31.05 & 26.61 & $-1.71$(0.06) & $-0.05$\\
J$001306.15+000431.9$ & 2.16 & 4829 & 6.7 & 13/3 & 1 & 0 & $<0.37$ &     0 & 1 & 31.17 & 31.23 & 26.44 & $-1.84$(0.06) & $-0.14$\\
J$002025.22+154054.6$ & 2.01 & 1595 & 19.9 & 560/200 & 1 & 0 & $3.20$ &     0 & 1 & 31.51 & 31.68 & 27.51 & $-1.60$(0.01) & $0.19$\\
J$002028.96+153435.8$ & 1.76 & 1595 & 19.9 & 100/36 & 1 & 1 & $<0.78$ &     0 & 0 & 30.89 & 31.06 & 26.85 & $-1.62$(0.02) & $0.04$\\
J$002155.52+001434.3$ & 1.83 & 0407030101 & 27.3 & 28/17 & 1 & 0 & $<0.93$ &     0 & 0 & 30.36 & 30.54 & 26.08 & $-1.71$(0.04) & $-0.17$\\
J$002308.74+002239.7$ & 2.05 & 0407030101 & 27.3 & 43/24 & 1 & 0 & $<0.64$ &     0 & 0 & 30.83 & 30.92 & 26.75 & $-1.60$(0.04) & $0.03$\\
J$002331.21-011045.6$ & 2.16 & 4079 & 1.9 & 2/2 & 0 & 0 & $<0.92$ &    17 & 0 & 30.14 & 30.68 & $<$26.76 & $<$$-1.50$ & $<$$0.07$\\
J$002825.59+003500.1$ & 1.97 & 4080 & 1.6 & 1/1 & 0 & 0 & $<0.88$ &     0 & 0 & 30.35 & 30.65 & $<$26.46 & $<$$-1.61$ & $<$$-0.04$\\
J$002917.31-002540.9$ & 1.77 & 0403160101 & 24.1 & 7/4 & 0 & 0 & $<0.81$ &     0 & 0 & 30.40 & 30.64 & $<$26.39 & $<$$-1.63$ & $<$$-0.06$\\
J$002954.96-001053.4$ & 2.35 & 0403160101 & 24.1 & 1/2 & 0 & 0 & $<0.91$ &     0 & 0 & 30.71 & 30.76 & $<$25.70 & $<$$-1.94$ & $<$$-0.35$\\
J$003131.44+003420.2$ & 1.74 & 2101 & 6.7 & 50/11 & 1 & 1 & $<0.44$ &     0 & 0 & 30.85 & 31.00 & 26.92 & $-1.57$(0.03) & $0.08$\\
J$003135.56+003421.3$ & 2.25 & 2101 & 6.7 & 32/10 & 1 & 0 & $<0.36$ &  3734 & 1 & 31.17 & 31.27 & 26.84 & $-1.70$(0.03) & $0.00$\\
J$003922.44+005951.7$ & 1.99 & 0203690101 & 40.6 & 61/32 & 1 & 0 & $<0.65$ &     0 & 0 & 30.45 & 30.89 & 26.71 & $-1.60$(0.03) & $0.02$\\
J$004206.18-091255.7$ & 1.78 & 4887 & 10.1 & 20/7 & 1 & 1 & $<0.65$ &     0 & 0 & 30.72 & 30.80 & 26.66 & $-1.59$(0.04) & $0.01$\\
J$004349.50+003930.2$ & 1.94 & 0090070201 & 20.3 & 40/11 & 1 & 0 & $<1.04$ &     0 & 0 & 30.30 & 30.48 & 27.07 & $-1.31$(0.04) & $0.22$\\
J$004526.26+143643.5$ & 1.96 & 6889 & 11.4 & 48/13 & 1 & 1 & $<0.77$ &     0 & 0 & 31.06 & 31.16 & 26.79 & $-1.68$(0.03) & $0.00$\\
J$004527.68+143816.1$ & 1.99 & 6889 & 11.4 & 7/5 & 1 & 0 & $<0.15$ &  6004 & 1 & 31.52 & 31.78 & 25.62 & $-2.37$(0.06) & $-0.55$\\
J$005018.84-005438.0$ & 2.33 & 4825 & 13.0 & 6/5 & 1 & 1 & $<0.78$ &     0 & 0 & 30.90 & 30.88 & 25.95 & $-1.89$(0.09) & $-0.27$\\
J$005025.13-005718.3$ & 2.55 & 4825 & 13.0 & 16/12 & 1 & 1 & $<0.69$ &     0 & 0 & 31.00 & 31.03 & 26.31 & $-1.81$(0.05) & $-0.16$\\
J$005102.42-010244.3$ & 1.88 & 4097 & 3.5 & 40/10 & 1 & 0 & $<0.02$ &     0 & 0 & 31.39 & 31.47 & 27.14 & $-1.66$(0.03) & $0.09$\\
J$005355.15-000309.3$ & 1.72 & 4830 & 7.1 & 18/10 & 1 & 0 & $<0.21$ &   550 & 3 & 30.81 & 31.22 & 26.16 & $-1.94$(0.04) & $-0.25$\\

\enddata
\tablenotetext{a}{The full version of this table is available in the electronic edition online.}
\tablenotetext{b}{The {\it Chandra} or {\it XMM-Newton} observation identification number.  Ten-digit numbers correspond to {\it XMM-Newton} observations, while shorter numbers correspond to {\it Chandra} observations.}
\tablenotetext{c}{The effective \mbox{X-ray} exposure reported by the CIAO or SAS toolchains for the extraction of the source region.}
\tablenotetext{d}{The observed-frame soft (0.5--2~keV) and hard (2--8~keV) counts in the source region.}
\tablenotetext{e}{A ``1'' indicates an X-ray detection; ``0'' indicates no detection.}
\tablenotetext{f}{A ``1'' indicates the source is in sample $B$; ``0'' indicates it is not in sample $B$.}
\tablenotetext{g}{The number of narrow \ion{Mg}{2} absorbtion systems found using the method described in \S\ref{fitSDSSSec}.}
\tablenotetext{h}{Monochromatic luminosities are given in units of erg~s$^{-1}$~Hz$^{-1}$.}
\tablenotetext{i}{The error, assumed to be dominated by the error on \mbox{X-ray} luminosity, is reported in parentheses.}
\end{deluxetable}
\clearpage
\end{landscape}

\clearpage
\begin{deluxetable}{lrrrrrr}
\tabletypesize{\scriptsize}
\tablecolumns{7}
\tablewidth{0pc}
\tablecaption{UV Emission Line Measurements\label{emInfoTab}\tablenotemark{a}}
\tablehead{\colhead{SDSS} & \colhead{\ion{C}{4} EW} & \colhead{\ion{C}{4} $\lambda$} & \colhead{\ion{C}{4} FWHM} & \colhead{\ion{C}{3}] EW} & \colhead{\ion{C}{3}] $\lambda$} & \colhead{\ion{C}{3}] FWHM} \\ \colhead{Source} & \colhead{(\AA)} & \colhead{(\AA)} & \colhead{(\AA)} & \colhead{(\AA)} & \colhead{(\AA)} & \colhead{(\AA)}}
\startdata
J$000654.10-001533.4$ & 51.8 & 1546.1 & 19.7 & 22.6 & 1904.1 & 25.7\\
J$000659.28-001740.8$ & 52.0 & 1555.3 & 20.5 & 13.9 & 1915.3 & 21.8\\
J$000717.73-002811.3$ & 27.6 & 1556.0 & 15.9 & 1.3 & 1918.1 &  1.4\\
J$001130.55+005550.7$ & 32.7 & 1542.6 & 37.2 & 15.8 & 1897.8 & 39.1\\
J$001247.12+001239.5$ & 37.1 & 1542.5 & 27.0 & 19.2 & 1909.0 & 28.6\\
J$001306.15+000431.9$ & 14.1 & 1539.2 & 33.5 & 13.9 & 1903.9 & 36.1\\
J$002025.22+154054.6$ & 31.8 & 1552.6 & 28.8 & 15.3 & 1910.8 & 36.0\\
J$002028.96+153435.8$ & 50.1 & 1544.3 & 28.9 & 27.2 & 1905.8 & 29.5\\
J$002155.52+001434.3$ & 19.5 & 1548.7 & 41.9 & 15.9 & 1905.0 & 34.0\\
J$002308.74+002239.7$ & 34.5 & 1545.4 & 16.1 & 10.0 & 1902.4 & 30.5\\
J$002331.21-011045.6$ & 20.8 & 1541.0 & 17.2 & 18.2 & 1896.1 & 42.1\\
J$002825.59+003500.1$ & 24.7 & 1552.1 & 25.0 & 30.4 & 1915.4 & 49.4\\
J$002917.31-002540.9$ & 80.1 & 1547.7 & 17.7 & 36.2 & 1907.9 & 22.4\\
J$002954.96-001053.4$ & 13.7 & 1544.8 & 28.8 & 12.9 & 1901.0 & 30.6\\
J$003131.44+003420.2$ & 59.6 & 1545.6 & 32.5 & 19.4 & 1901.2 & 47.7\\
J$003135.56+003421.3$ & 46.9 & 1537.1 & 22.1 & 36.3 & 1895.4 & 34.0\\
J$003922.44+005951.7$ & 31.3 & 1551.4 & 13.0 & 6.3 & 1912.5 & 23.9\\
J$004206.18-091255.7$ & 32.5 & 1546.6 & 22.3 & 22.2 & 1904.8 & 29.5\\
J$004349.50+003930.2$ & 87.0 & 1545.6 & 33.2 & 41.8 & 1898.3 & 55.8\\
J$004526.26+143643.5$ & 39.9 & 1544.1 & 29.0 & 17.0 & 1906.0 & 41.8\\
J$004527.68+143816.1$ & 14.3 & 1547.7 & 19.9 & 19.0 & 1896.4 & 50.5\\
J$005018.84-005438.0$ & 14.4 & 1545.2 & 31.2 & 14.5 & 1906.4 & 41.0\\
J$005025.13-005718.3$ & 42.4 & 1555.2 & 21.7 & 28.4 & 1918.7 & 32.6\\
J$005102.42-010244.3$ & 29.5 & 1546.3 & 36.2 & 21.5 & 1909.1 & 43.5\\
J$005355.15-000309.3$ & 23.7 & 1539.2 & 27.6 & 18.4 & 1899.8 & 39.0\\

\enddata
\tablenotetext{a}{The full version of this table is available in the electronic edition online.}
\end{deluxetable}

\clearpage

\begin{deluxetable}{lrrrrr}
\tablecolumns{3}
\tablewidth{0pc}
\tablecaption{Sample Information\label{sampleInfoTab}}
\tablehead{\colhead{Sample} & \colhead{Total} & \colhead{Radio-Quiet} & \colhead{Radio-Loud} & \colhead{Radio-Quiet} & \colhead{Radio-Loud} \\ \colhead{} & \colhead{Size} & \colhead{Non-BAL QSOs\tablenotemark{a}} & \colhead{Non-BAL QSOs\tablenotemark{a}} & \colhead{BAL QSOs\tablenotemark{a}} & \colhead{BAL QSOs\tablenotemark{a}}}
\startdata
$A$ & 536 & 433/354 & 34/32 & 64/43 & 5/5\\
$B$ & 163 & 139/139 & 10/10 & 13/11 & 1/1\\

\enddata
\tablenotetext{a}{The number of sources is reported as ``$a$/$b$,'' where $a$ is the total number of sources of the specified type, and $b$ is the number of those sources which were detected in \mbox{X-rays}.}
\end{deluxetable}

\begin{deluxetable}{lrrllrrr}
\tablecolumns{3}
\tablewidth{0pc}
\tablecaption{Tests for Emission Line Correlations with $\Delta\alpha_{OX}$\label{lineCorrTestsTab}}
\tablehead{\colhead{Relation} & \colhead{Confidence} & \colhead{Confidence} \\ \colhead{} & \colhead{BAL QSOs Included} & \colhead{Only Non-BAL QSOs}}
\startdata
\ion{C}{4} EW & $>$99.99 & $>$99.99\\
\ion{C}{4} Wavelength & 99.2 & 97.0\\
\ion{C}{4} FWHM & 67.5 & 32.0\\
\ion{C}{3}] EW & 94.8 & 85.2\\
\ion{C}{3}] Wavelength & 31.8 & 36.2\\
\ion{C}{3}] FWHM & $>$99.99 & 99.96\\
\ion{C}{4} - \ion{C}{3}] Wavelength & 90.3 & 95.8\\

\enddata
\end{deluxetable}

\end{document}